\documentclass{article}
\usepackage{graphicx}
\usepackage{amsfonts}
\usepackage{amssymb}
\usepackage{latexsym}

\newcommand{\squeezelist}{\setlength{\itemsep}{0pt}}

\def\e{{\epsilon}}
\def\b{{\beta}}
\def\a{{\alpha}}
\def\G{{\Gamma}}
\def\g{{\gamma}}
\def\t{{\theta}}
\def\l{{\lambda}}
\def\bP{{\partial P}}
\def\bQ{{\partial Q}}
\def\T{{\cal T}}
\newenvironment{pf}{\unskip{\bf Proof:}}{\unskip{\hfill $\Box$}}

\newcommand{\lemlab}[1]{\label{lemma:#1}}
\newcommand{\theolab}[1]{\label{theo:#1}}

\newcommand{\eqlab}[1]{\label{eq:#1}}

\newcommand{\factlab}[1]{\label{fact:#1}}
\newcommand{\tablab}[1]{\label{tab:#1}}
\newcommand{\figlab}[1]{\label{fig:#1}}
\newcommand{\seclab}[1]{\label{section:#1}}

\newcommand{\lemref}[1]{\ref{lemma:#1}}

\newcommand{\theoref}[1]{\ref{theo:#1}}

\newcommand{\factref}[1]{\ref{fact:#1}}
\newcommand{\tabref}[1]{\ref{tab:#1}}
\newcommand{\figref}[1]{\ref{fig:#1}}
\newcommand{\eqref}[1]{(\ref{eq:#1})}
\newcommand{\secref}[1]{\ref{section:#1}}

\newtheorem{theorem}{Theorem}[section]
\newtheorem{lemma}[theorem]{Lemma}
\newtheorem{cor}[theorem]{Corollary}
\newtheorem{conj}{Conjecture}[section]
\newtheorem{fact}{Fact}[section]
{\catcode`\@=11
\gdef\setft#1#2#3{%
\def\@oddfoot{
{\setbox0=\hbox{#1}
\setbox1=\hbox{#3}
\ifdim\wd0>\wd1
\dimen0=\wd0
\box0\hfil#2\hfil\hbox to\dimen0{\hfil\hfil\box1}
\else \dimen0=\wd1
\hbox to\dimen0{\box0\hfil }\hfil#2\hfil\box1 \fi
}}} }

\def\complaint#1{}
\def\withcomplaints{
\newcounter{mycomplaints}
\def\complaint##1{\refstepcounter{mycomplaints}%
\ifhmode%
\unskip%
{\dimen1=\baselineskip \divide\dimen1 by 2 %
\raise\dimen1\llap{\tiny -\themycomplaints-}}\fi%
\marginpar{\tiny [\themycomplaints]: ##1}}%
}

\title{\bf Examples, Counterexamples, and \\
Enumeration Results for\\
Foldings and Unfoldings\\
between Polygons and Polytopes}
\author{%
Erik Demaine \and Martin Demaine \and Anna Lubiw\thanks{
Dept.\ Comput\ Sci., Univ. of Waterloo,
Waterloo, Ontario N2L 3G1, Canada.
\texttt{\{eddemaine,mdemaine,alubiw\}@\penalty \exhyphenpenalty uwaterloo.ca}.
}
\\ \and
Joseph~O'Rourke\thanks{
Dept.\ Comput.\ Sci., Smith Col\-lege, North\-ampton, 
MA 01063, USA.
\texttt{orourke@\penalty \exhyphenpenalty cs.smith.edu}.
Supported by NSF grant CCR-9731804.}
}
\begin{document}
\maketitle
\begin{abstract}
We investigate how to make the surface of a convex polyhedron 
(a {\em polytope\/}) by folding up a
polygon and gluing its perimeter shut, and the reverse process
of cutting open a polytope and unfolding it to a polygon.
We explore basic enumeration questions in both directions:
Given a polygon, how many foldings are there?
Given a polytope, how many unfoldings are there to simple polygons?
Throughout we give special attention to convex polygons,
and to regular polygons.
We show that every convex polygon folds to an infinite number of
distinct polytopes, but that
their number of combinatorially distinct gluings is polynomial.  
There are, however, simple polygons with an exponential number
of distinct gluings.

In the reverse direction, we show that there are polytopes with an
exponential number of distinct cuttings that lead to simple unfoldings.
We establish necessary conditions for a polytope to have convex unfoldings,
implying, for example,
that among the Platonic solids, only the tetrahedron has a convex
unfolding.
We provide an inventory of the polytopes that may unfold to regular polygons,
showing that, for $n>6$, there is essentially only one
class of such polytopes.
\end{abstract}

\newpage
\tableofcontents
\newpage
\pagenumbering{arabic}
\setcounter{page}{1}

\section{Introduction}
\seclab{Introduction}
We explore the process of folding a simple polygon 
by gluing its perimeter shut to form
a convex polyhedron, and its reverse, cutting a
convex polyhedron open and flattening its surface to a
simple polygon.
We restrict attention to convex polyhedra (henceforth,
{\em polytopes\/}), 
and to simple (i.e., nonself-intersecting,
nonoverlapping) polygons (henceforth just {\em polygons}).
The restriction to nonoverlapping polygons is natural,
as this is important to the manufacturing
applications~\cite{o-fucg-00}.
The restriction to convex polyhedra is made primarily to
reduce the scope of the problem.
See ~\cite{bddloorw-uscop-98} and~\cite{bdek-up-99}
for a start on unfolding nonconvex polyhedra.

Much recent work on unfolding revolves around an 
open problem that seems to have been first mentioned
in print in~\cite{s-cpcn-75} but is probably much older:
Can every polytope be cut along edges and unfolded
flat to a (simple) polygon?
Cutting along edges leads to {\em edge unfoldings\/};
we will not follow this restriction here.
Thus our work is only indirectly related to this edge-unfolding
question.

In some sense this report is a continuation
of the investigation started in~\cite{lo-wcpfp-96},
which detailed an $O(n^2)$ algorithm
for deciding when a polygon may be folded to a polytope,
with the restriction that each edge of the polyon perimeter
glues to another complete edge:
{\em edge-to-edge gluing}.
But here we do not following this restriction,
permitting arbitrary perimeter gluings.
Moreover, we do not consider algorithmic questions.
Rather we concentrate on enumerating the number of foldings
and unfoldings between polygons and polytopes.
We pay special attention to convex polygons;
following Shephard~\cite{s-cpcn-75}, we call an
unfolding of a polytope that produces a convex polygon
a {\em convex unfolding}.
Within the class of polytopes, we sometimes
use the five regular polytopes as examples;
within the class of convex polygons, we additionally
focus on regular polygons.

The basic questions we ask are:
\begin{enumerate}
\squeezelist
\item How many combinatorially different foldings of a polygon lead to a polytope?
\item How many geometrically different polytopes may be folded from one polygon?
\item How many combinatorially different cuttings of a polytope lead to polygon unfoldings?
\item How many geometrically different polygons may be unfolded from one polytope?
\end{enumerate}
Our answers to these four
questions are crudely summarized in Table~\tabref{results}, 
whose four rows correspond to
the four questions above, and whose columns are for general, convex, and
regular polygons.
We will not explain the entries in the table here, but only remark
that the increased constraints provided by convex and regular
polygons reduces the number of possibilities.
\begin{table}[htbp]
\begin{center}\begin{tabular}{| l | l | c | c | c |}
        \hline

\mbox{}
	& \mbox{}
	& General
	& Convex 
	& Regular 
	\\
\mbox{}
	& \mbox{}
	& \mbox{}
	& Polygons
	& Polygons
        \\ \hline \hline
	
Foldings
        & gluing trees
	& $2^{\Omega(n)}$, $O(n^{2\l-2})$
	& $O(n^3)$
	& $O(1)$
        \\ \cline{2-5}
\mbox{}
        & polytopes
	& $\infty$
	& $\infty$
	& $2$ classes
	\\ \hline
	
Unfoldings
        & cut trees
	& $2^{\Omega(n)}$, $2^{O(n^2)}$
	& ?
	& $O(1)$
        \\ \cline{2-5}
\mbox{}
        & polygons
	& $\infty$
	& $0$, $\infty$
	& $O(1)$
	\\ \hline
\end{tabular}
\tablab{results}
\caption{Summary of Results.  $n$ is the number of polygon vertices
or polytope vertices; $\l$ is the number of leaves of the gluing tree;
the symbol $\infty$ represents nondenumerably infinite, i.e., a
continuum.}
\end{center}
\end{table}

A key tool in our work is
a powerful theorem of
Aleksandrov, which we describe and
immediately apply in Section~\secref{Aleksandrov}.
We then define the two main combinatorial objects
we study, cut trees and gluing trees, and make
clear exactly how we count them.
We then explore constraints on convex unfoldings
in Section~\secref{Convex.Unfoldings}
before proceeding to the general enumeration bounds
in Table~\tabref{results}
in Sections~\secref{Counting.Gluings}-\secref{Noncongruent.Polygons}.
A final section (\secref{Regular}) concentrates on regular polygons

\section{Aleksandrov's Theorem}
\seclab{Aleksandrov}

Aleksandrov proved a far-reaching generalization of
Cauchy's rigidity theorem
in~\cite{a-kp-58} that gives simple conditions for any folding
to a polytope.
Let $P$ be a polygon and $\bP$ its boundary.
A {\em gluing\/} 
maps $\bP$ to $\bP$ in a length-preserving
manner, as follows.
$\bP$ is partitioned by a finite number of distinct points
into a collection of open intervals whose closure covers $\bP$.
Each interval is mapped one-to-one (i.e., {\em glued\/})
to another interval of equal length.
Corresponding endpoints of glued intervals are glued
together (i.e., identified).
Finally, gluing is considered
transitive: if points $a$ and $b$ glue to point $c$, then
$a$ glues to $b$.%
\footnote{
        What we call {\em gluing\/} is sometimes called
        {\em pasting\/}~\cite[p.~13]{az-igs-67}.
	In the theory of complexes, it is sometimes called
	{\em topological identification}~\cite[p.~116]{h-cit-79}.
}
Aleksandrov proved that any gluing that satisfies these two conditions 
corresponds to a unique polytope:
\begin{enumerate}
\item No more than $2 \pi$ total face angle is glued together at any point; and
\item The complex resulting from the gluing is homeomorphic to a sphere.
(This condition is satisfied if, when
$\bP$ is viewed as a topological circle, and the interval
gluings as chords of the circle, then no pair of chords cross in the
$\bP$-circle.)
\end{enumerate}
Aleksandrov calls any complex (not necessarily a single polygon)
that satisfies these properties a {\em net}~\cite{a-kp-58}.%
\footnote{
	This may derive from the German translation, {\em Netz}.
	In fact, the Russian word Aleksandrov used is
	closer to ``unfolding.''
}
We call a gluing that satisfies these conditions
an {\em Aleksandrov gluing}.

Although an
Aleksandrov gluing of a polygon forms a unique polytope, 
it is an open problem to compute
the three-dimensional structure of the polytope~\cite{o-fucg-00}.
Note that there is no specification of the fold (or ``crease'') lines;
and yet they are uniquely determined.
Henceforth we will say a polygon {\em folds\/} to
a polytope whenever it has an Aleksandrov gluing.

We should mention two features of Aleksandrov's theorem.
First, the polytope whose existence is guaranteed may be
{\em flat}, that is, a doubly-covered convex polygon.
We use the term ``polytope'' to include flat polyhedra.
Second,
condition~(2) specifies a face angle $\le 2 \pi$.
The case of equality with $2 \pi$ leads to a point
on the polytope at which there is no curvature,
i.e., a nonvertex.  We make explicit 
what counts as a vertex below.

\paragraph{Polygon/Polytope Notation.}
We will use $P$ throughout the paper for a polygon,
and $Q$ for a polytope.
Their boundaries are $\bP$ and $\bQ$ respectively.
The {\em curvature\/} $\g(x)$ of a point $x \in \bQ$
is $2 \pi$ minus the
sum of the face angles incident to $x$.
This ``angle deficit'' corresponds to the notion of Gaussian curvature.
We define vertices of polygons and polytopes to be
{\em essential\/} in the sense
that the boundary is not flat there:
the interior angle at a polygon vertex is different from
$\pi$, and
the curvature at a polytope vertex is different from $0$.
Because of these definitions, there is no direct correspondence
between the vertices of a polytope $Q$ and the vertices of
a polygon $P$ unfolding of $Q$:  a vertex of $Q$ may or may
not unfold to a vertex of $P$; and a vertex of $P$ may or
may not fold to a vertex of $Q$
(see Section~\secref{cut.glue.comp}).
At the risk of confusion, we will use the terms
``vertex'' and ``edge'' for both polygons and polytopes,
but reserve ``node'' and ``arc'' for graphs.
We will use $n$ for the number of vertices of $P$ or $Q$, letting
the context determine which.

We will also freely employ two types of paths on the
surface of a polytope:
{\em geodesics}, which unfold (or ``develop'') to straight lines,
and {\em shortest paths}, geodesics which are in addition
shortest paths between their endpoints.
See, e.g., \cite{aaos-supa-97} for details and basic properties.

\subsection{Perimeter Halving}
As a straightforward application of Aleksandrov's theorem, we
prove that every convex polygon folds to a polytope.
We will see in Section~\secref{Sharp.Vertices} that the converse does not hold.

For two points $x,y \in \bP$, define $(x,y)$ be the open interval
of $\bP$ counterclockwise from $x$ to $y$,
and let $|x,y|$ be its length.
Define a {\em perimeter-halving gluing\/}
as one which glues $(x,y)$ to $(y,x)$.

\begin{lemma}
Every convex polygon folds to a polytope via perimeter halving.
\lemlab{perim.halving}
\end{lemma}
\begin{pf}
Let the perimeter of a convex polygon $P$ be $L$.
Let $x \in \bP$ be an arbitrary point on the boundary of $P$,
and let $y \in \bP$ be the midpoint of perimeter around $\bP$
measured from $x$, i.e., $y$ is the unique point
satisfying $|x,y|= |y,x| = L/2$.
See Fig.~\figref{perim.halving} for an example.
\begin{figure}[htbp]
\centering
\includegraphics[scale=1.0]{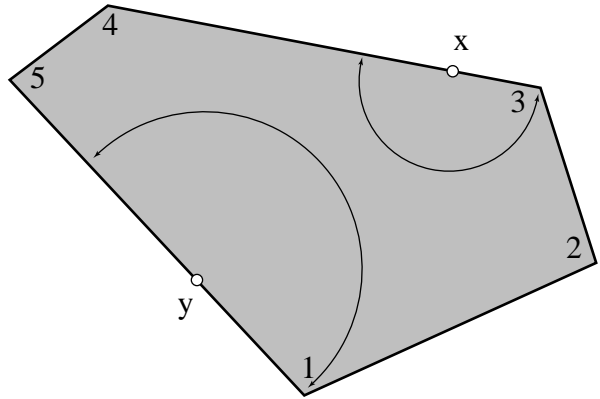}
\caption{A perimeter-halving fold of a pentagon. The gluing mappings
of vertices $v_1$ and $v_3$ are shown.}
\figlab{perim.halving}
\end{figure}

Now glue $(x,y($ to $(y,x)$ in the natural way, mapping each
point $z$ with $|x,z|=d$ to the point $z'$ the same distance
from $x$ in the other direction: $|z',x|=d$.
We claim this is an Aleksandrov gluing.  It is a gluing
by construction.  Because $P$ is convex, each point along
the gluing path has $\le 2 \pi$ angle incident to it:
the gluing of two nonvertex points results in exactly $2\pi$,
and if either point is a vertex, the total angle is strictly
less than $2\pi$.
The resulting surface is clearly homeomorphic to a sphere.
By Aleksandrov's theorem, this gluing corresponds to a unique
polytope $Q_x$.
\end{pf}

In an Aleksandrov gluing of a polygon, a point in the interior of a
polygon edge that glues only to itself, i.e., where a crease folds
the edge in two, is called a {\em fold point}.  
A fold point corresponds
to a leaf of the gluing tree, and becomes a vertex of the polytope
with curvature $\pi$.
Points $x$ and $y$ in the above proof are fold points.
In Theorem~\theoref{continuum} we will show that different choices
of $x$ result in distinct polytopes $Q_x$, leading to the
conclusion that every convex polygon folds to an infinite number
of polytopes.

\section{Cut Trees and Gluing Trees}
\seclab{cut.glue.trees}

The four main objects we study are polygons, polytopes,
cut trees, and gluing trees.
It will be useful in spots to distinguish between 
a {\em geometric\/} tree $\T$ composed of a union of line
segments, and the more familiar {\em combinatorial\/} tree $T$
of nodes and arcs.
A {\em geometric cut tree\/} $\T_C$ for a polytope $Q$ is a tree drawn on $\bQ$,
with each arc a polygonal path, 
which leads to a polygon unfolding when the surface is cut along $\T$,
i.e., flattening $Q \setminus \T$ to a plane.
A {\em geometric gluing tree\/} $\T_G$ specifies how $\bP$ is glued to
itself to fold to a polytope.
There is clearly a close correspondence between $\T_C$ and $\T_G$,
which are in some sense the same object, one viewed from
the perspective of unfolding, one from the perspective of folding.
It will nevertheless be useful to retain a distinction between
them, and especially their combinatorial counterparts,
which we define below after stating some basic properties.

\subsection{Cut Trees}
\seclab{Cut.Trees}

\begin{lemma}
If a polygon $P$ folds to a polytope $Q$, $\bP$ maps
to a tree $\T_C \subset \bQ$, the geometric cut tree, with
the following properties:
\begin{enumerate}
\item $\T_C$ is a tree.
\item $\T_C$ spans the vertices of $Q$.
\item Every leaf of $\T_C$ is at a vertex of $Q$.
\item 
A point of $\T_C$ of
degree $d$ (i.e., one with $d$ incident segments)
corresponds to exactly $d$ points of $\bP$.
Thus a leaf corresponds to a unique point
of $\bP$.
\item Each arc of $T_C$ is a polygonal path on $Q$.
\end{enumerate}
\lemlab{cut.tree}
\end{lemma}
\begin{pf}
\begin{enumerate}
\item If $\T_C$ contained a cycle, then it would unfold to disconnected pieces,
contradicting the assumption that $Q$ is folded from a single polygon $P$.
Thus $\T_C$ is a forest.  But because $\T_C$ is constructed by gluing
the connected path $\bP$ to itself, it must be connected.
So $\T_C$ is a tree.
\item If a vertex $v$ of $Q$ is not touched by $\T_C$, then,
because $Q$ is not flat at $v$, $P$ is not planar,
a contradiction to the assumption that $P$ is a polygon.
\item Suppose a leaf $x$ of $\T_C$ is interior to a face or edge of $Q$.
Then it is surrounded by $2 \pi$ face angle on $Q$, and so unfolds to a
point $x$ of $P$ similarly surrounded.  But by assumption, $x$ is on the
boundary of a simple polygon $P$, a contradiction.
\item Gluing exactly two distinct points of $x,y \in \bP$ together implies that
neighborhoods of $x$ and $y$ are glued, which leads to the interior of
an arc of the cut tree, i.e., a degree-$2$ point of $\T_C$.
Note that either or both of these points might be vertices of $P$.
In general, if $p \in \T_C$ has $d$ incident cut segments,
$p$ unfolds to $d$ distinct points of $\bP$.
\item If an arc of $\T_C$ is not a polygonal path,
then neither side unfolds to a polygonal path, contradicting
the assumption that $P$ is a polygon.
\end{enumerate}
\end{pf}

When counting cut trees, we will rely on their combinatorial
structure.
There are several natural definitions of this structure, which
are useful in different circumstances.  We first discuss some
of the options.
\begin{enumerate}
\item Make every segment of $\T_C$ an arc of $T_C$.
Although this is very natural, it means there are an infinite number
of different cut trees for any polytope, for the path
between any two polytope vertices could be an arbitrarily
complicated polygonal path, leading to different combinatorial
trees.
\item Make every point where a path of $\T_C$ crosses an edge
of the polytope a node of $T_C$.  This again leads to trivially
infinite numbers of cut trees when a path of $\T_C$ zigzags 
back and forth over an edge of $Q$. 
\item Exclude this possibility by forcing the paths between
polytope vertices to be
geodesics, and again make polytope edge crossings nodes of $T_C$.
This excludes many interesting cut trees---all those where a
polygon vertex is glued to a point with angle sum $2\pi$.
\item Make every maximal path of $\T_C$ consisting only of
degree-$2$ points a single arc of $T_C$.  This has the
undesirable effect of having polytope vertices in the interior
of such a path disappear from $T_C$.
\end{enumerate}

Threading between these possibilities, we define
the {\em combinatorial cut tree\/} $T_C$ corresponding
to a geometric cut tree $\T_C$ as
the labeled graph with a node (not necessarily labeled)
for each point of $\T_C$ with degree not equal to $2$,
and a labeled node for each point of $\T_C$ that corresponds
to a vertex of $Q$ (labeled by the vertex label);
arcs are determined by the polygonal paths of $\T_C$
connecting these nodes.
An example is shown in
Fig.~\figref{cut.tree}.  Note that not every node 
of the tree is labeled, but
every polytope vertex label is used at some node.
All degree-$2$ nodes are labeled.
\begin{figure}[htbp]
\centering
\includegraphics[width=0.6\linewidth]{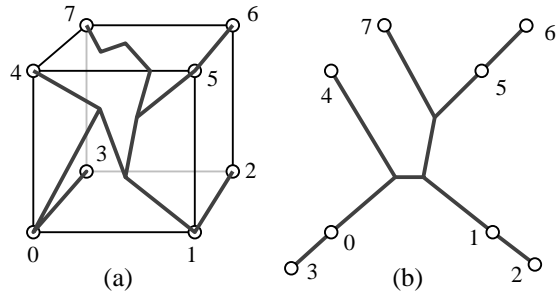}
\caption{(a) Geometric cut tree $\T_C$ on the surface of a cube; 
(b) The corresponding combinatorial cut tree $T_C$.}
\figlab{cut.tree}
\end{figure}

Although this definition avoids some of the listed pitfalls,
it does have the
undesirable consequence of counting different geodesics
on $\bQ$ between two polytope vertices as the same arc of $T_C$.
Thus 
the two unfoldings shown in Fig.~\figref{not.sp} (below)
have the same combinatorial cut tree under our definition,
even though the geodesic in (c) spirals twice around
compared to once around in (a).

\subsection{Gluing Trees}
Let a convex polygon $P$ have vertices $v_1,\ldots,v_n$,
labeled counterclockwise,
and edge $e_i$, $i=1,\ldots,n$ the open segment of $\bP$
after $v_i$.
There is less need to discuss the geometric gluing tree,
so we concentrate on the combinatorial gluing tree $T_G$.
$T_G$ is a tree representing the
identification of $\bP$ with itself.
Any point of $\bP$ that is identified with more or less than
one other distinct point of $\bP$ becomes a node
of $T_G$, as well as any point to which a vertex is
glued.
(Note that this means there may be nodes of degree $2$.)
So every vertex of $P$ maps to a node of $T_G$;
each node is labeled with the set of all the
elements (vertices or edges) that are glued together there.
A leaf that is a fold point is labeled by the edge label only.
Every nonleaf node has at least one vertex label, and at most one edge label.
A simple example is shown in Fig.~\figref{tri.tetra}.%
\footnote{
	Gluing trees can be drawn by folding up the polygon toward
	the viewer (as in this figure), or folding the polygon
	away.  We employ both conventions but always note which
	is followed.
}
Here the central node of $T_G$ is assigned
the label $\{v_1,v_3,e_3\}$.
\begin{figure}[htbp]
\centering
\includegraphics[width=10cm]{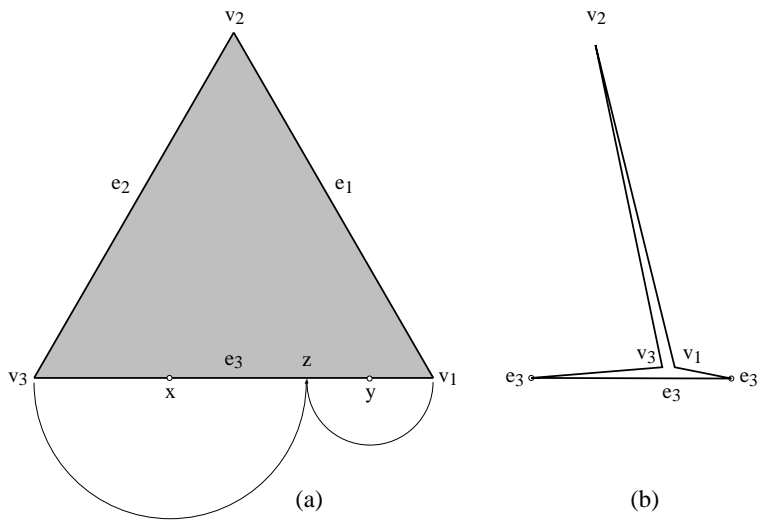}
\caption{(a) A gluing of an equilateral triangle $P$: $v_1$ and $v_3$
are glued to point $z$;
(b) the corresponding gluing tree $T_G$ [folding up].
Points $x$ and $y$ become fold points of the resulting tetrahedron.}
\figlab{tri.tetra}
\end{figure}
A more complicated example is shown in 
Fig.~\figref{zed.tetra2}.\footnote{
	We found this example by an enumeration algorithm that
	will not be discussed in this report.
}
The polygon shown folds (amazingly!) to a tetrahedron
by creasing as illustrated in (a).  All four tetrahedron
vertices are fold points.  The corresponding gluing tree
is shown in (b) of the figure.  The two interior nodes
of $T_G$ have labels $\{v_1,v_6,e_1\}$ and $\{v_2,v_5,e_5\}$.
\begin{figure}[htbp]
\centering
\includegraphics[width=\linewidth]{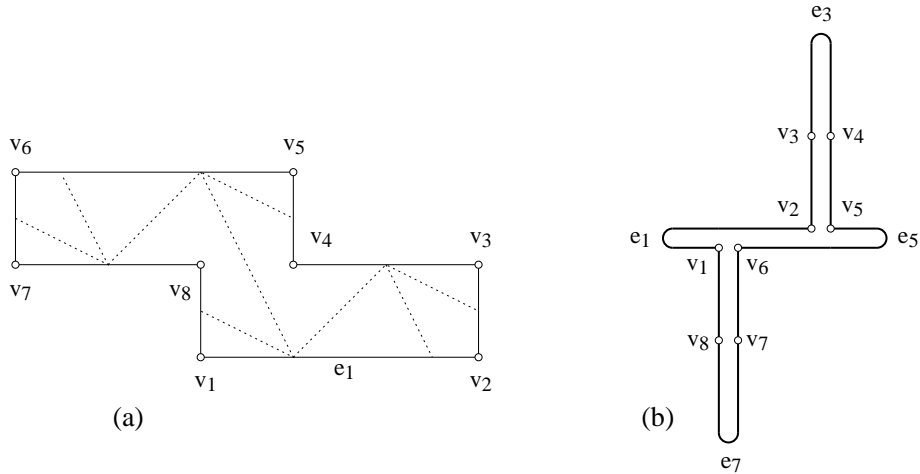}
\caption{(a) A polygon, with fold creases shown dotted;
(b) A gluing tree $T_G$ [folding away] corresponding to the crease pattern.}
\figlab{zed.tetra2}
\end{figure}

Later (Lemma~\lemref{gluing.tree}) will show that the gluing tree
is determined by a relatively sparse set of gluing instructions.

\subsection{Comparison of Cut and Gluing Trees}
\seclab{cut.glue.comp}

\begin{lemma}
Let $T_C$ be a combinatorial cut tree for polytope $Q$
that unfolds to a polygon $P$, and let $T_G$ be the 
combinatorial gluing
tree that folds $P$ to $Q$.
If all degree-$2$ nodes are removed by contraction,
$T_C$ and $T_G$ are isomorphic as unlabeled graphs.
\lemlab{cut.glue}
\end{lemma}
\begin{pf}
Let $(a,b,c)$ be three consecutive nodes on a path in a tree $T$,
with $b$ of degree $2$. Removing $b$ by contraction deletes $b$
and replaces it with the arc $(a,c)$.
Applying this to both $T_C$ and $T_G$ produces two trees
$T'_C$ and $T'_G$ without degree-$2$ nodes.  As the trees were
defined to include nodes for each point whose degree differs
from $2$, it must be that $T'_C$ and $T'_G$ have isomorphic
structures.  Of course they are labeled differently, but without
the labels, they are isomorphic graphs.
\end{pf}

Note that vertices in $Q$ and vertices in $P$ do not necessarily
map to one another:  A vertex of $Q$ can map to an 
interior point of $\bP$, and a vertex of $P$ can map to
a point interior to a face or edge of $Q$.
This affects the labeling of the two trees, but
they have essentially the same structure.

\section{Cut Trees for Convex Unfoldings}
\seclab{Convex.Unfoldings}
Before embarking on general enumeration results, 
we specialize the discussion to convex unfoldings,
and derive some constraints on the possible cut trees that
lead to convex unfoldings.

\subsection{Stronger Characterization}
We now sharpen the characterization of cut trees
(and via Lemma~\lemref{cut.glue}, of gluing trees)
under the restriction that the unfolding must be a convex 
polygon.
We first strengthen Lemma~\lemref{cut.tree}(5), which
only required arcs to be polygonal paths:
\begin{lemma}
Every arc of
a cut tree $T_C$ that leads to a convex unfolding
must be a {\em geodesic\/} on $Q$ (paths that unfold to straight segments),
but arcs might not be
shortest paths on $Q$.
\lemlab{not.sp}
\end{lemma}
\begin{pf}
Suppose an arc $a$ of $T$ is not a geodesic.  Then it does
not unfold to a straight line. Suppose a point $x \in a$ is a point
in the relative interior of $a$
at which the unfolding is locally not straight.  Then only one
of the two points of $\bP$ that correspond to $x$ can have an
interior angle $\le \pi$ in $P$, showing that $P$ has at least one
reflex angle.
This establishes that arcs of $T_C$ must be geodesics.
We now show that this claim cannot be strengthened to shortest
paths by an explicit example.

Let $Q$ be a doubly-covered rectangle with
vertices $v_i$, $i=1,2,3,4$, as shown in Fig.~\figref{not.sp}(a).
Let $x$ be the midpoint of edge $v_1 v_4$.
Let $T_C$ be the path $(v_1, v_2, x, v_3, v_4)$, where the subpath
$(v_2, x, v_3)$ is half on the upper rectangular face,
and half on the bottom face.  Clearly this subpath is not a shortest path,
although it is a geodesic.  The corresponding convex unfolding
is shown in Fig.~\figref{not.sp}(b).
\begin{figure}[htbp]
\centering
\includegraphics[width=0.7\linewidth]{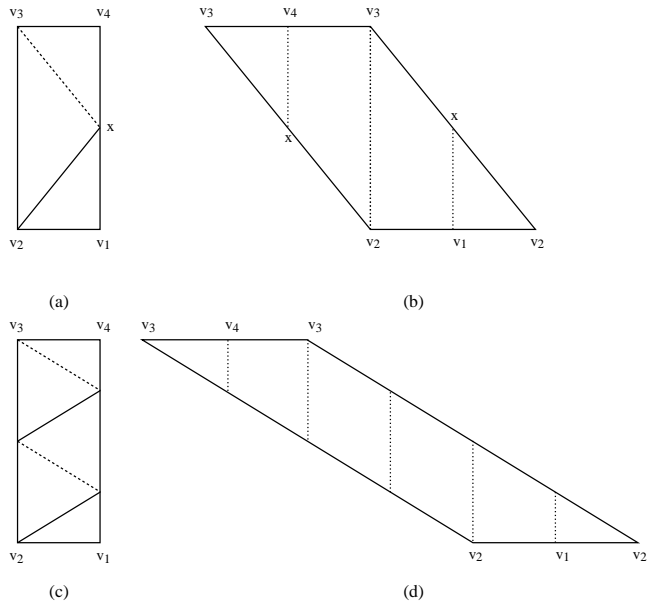}
\caption{(a)Doubly covered rectangle with cut path; (b) Unfolding.
(c-d): Another cut path and its unfolding.}
\figlab{not.sp}
\end{figure}

This example can be modified to a nondegenerate 
``sliver'' tetrahedron by perturbing
one vertex to lie slightly out of the plane of the other three.
\end{pf}

\noindent
Fig.~\figref{not.sp}(c-d) shows that we cannot even bound the
length of a geodesic arc of $T_C$.

One immediate corollary of Lemma~\lemref{not.sp} is that cuts need
not follow polytope edges (which are all shortest paths),
i.e., not every convex unfolding is an edge unfolding.

\subsection{Necessary Conditions: Sharp Vertices}
\seclab{Sharp.Vertices}

We define a vertex of a polytope to be {\em sharp\/} if it has curvature $\ge \pi$,
and {\em round\/} if its curvature is $< \pi$.
The following theorem gives a simple necessary condition for a polytope
to have a convex unfolding.
We employ this fact implied by the Gauss-Bonnet theorem:

\begin{fact}
The sum of the curvatures of all the vertices of a polytope is
exactly $4 \pi$.
\factlab{4pi}
\end{fact}

\begin{theorem}
If a polytope $Q$ has a convex unfolding via a cut
tree $T_C$, then each leaf of $T_C$ is at a sharp vertex.
Moreover, $Q$ must
have at least two sharp vertices.
\theolab{two.sharp}
\end{theorem}
\begin{pf}
Let $P$ be a convex polygon to which $Q$ unfolds via cut tree $T_C$.
By Lemma~\lemref{cut.tree}(3), the leaves of $T_C$ are at vertices
of $Q$.  Let $x$ be a leaf of $T_C$, at a vertex $v$ with curvature 
$\g(v) = \g$.
Point $x \in \bQ$ corresponds to a unique point $y \in \bP$
by Lemma~\lemref{cut.tree}(4).
The internal angle at $y$ in $P$ is $2 \pi - \g$.  Because
$P$ is convex, we must have 
$$
2 \pi - \g \le \pi
$$
and so $\g(v) \ge \pi$.  Thus $v$ is sharp.
Because $T_C$ must have at least two distinct leaves, the lemma follows.
\end{pf}

\begin{cor}
Of the five Platonic solids, only the regular tetrahedron has
a convex unfolding.
\end{cor}
\begin{pf}
The curvatures at the vertices of the solids are:
\begin{eqnarray*}
2 \pi - 3(\pi/3)  = &  \pi & \mbox{} \\
2 \pi - 3(\pi/2)  = &  \pi/2 & < \pi \\
2 \pi - 4(\pi/3)  = &  2\pi/3 & < \pi \\
2 \pi - 3(3\pi/5) = &  \pi/5 & < \pi \\
2 \pi - 5(\pi/3)  = &  \pi/3 & < \pi
\end{eqnarray*}
Only the tetrahedron has sharp vertices.
\end{pf}

We next show that two natural extensions of the previous results
fail.

\begin{lemma}
There is a tetrahedron with no convex unfolding.
\lemlab{not.tetra}
\end{lemma}
\begin{pf}
Let $Q_1$ be a tetrahedron whose vertices $v_1,v_2,v_3$ form an
equilateral triangle base in the $xy$-plane, with apex $v_4$
centered at a great height $z$ above. 
See Fig.~\figref{not.tetra}. 
Let $\g_i$ be the curvature of
vertex $v_i$.
If the face angle of each triangle incident to $v_4$ is $\e$,
then $\g_4 = 2 \pi - 3 \e$,
and $\g_i$ for $i=1,2,3$ is
$$2 \pi - [\pi/3 + 2(\pi-\e/2)] = 2 \pi /3 + \e$$
Choosing $z$ large makes $\e$ small,
and then $Q_1$ has just one sharp vertex.
Theorem~\theoref{two.sharp} then establishes the claim.
\end{pf}

\begin{figure}[htbp]
\begin{minipage}[b]{0.45\linewidth}
\centering
\includegraphics[scale=1.0]{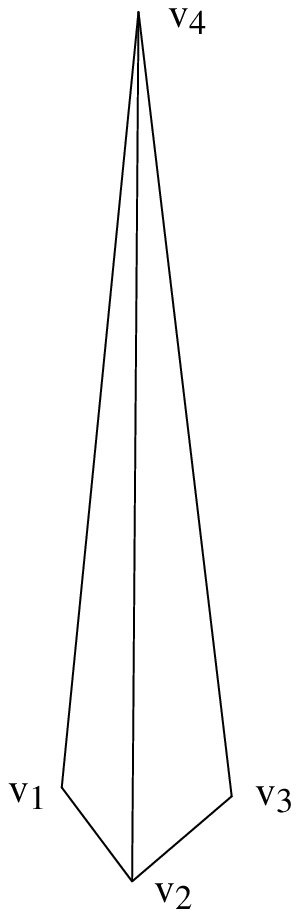}
\caption{A tetrahedron $Q_1$ without a convex unfolding.}
\figlab{not.tetra}
\end{minipage}%
\hspace{5mm}%
\begin{minipage}[b]{0.45\linewidth}
\centering
\includegraphics[scale=1.0]{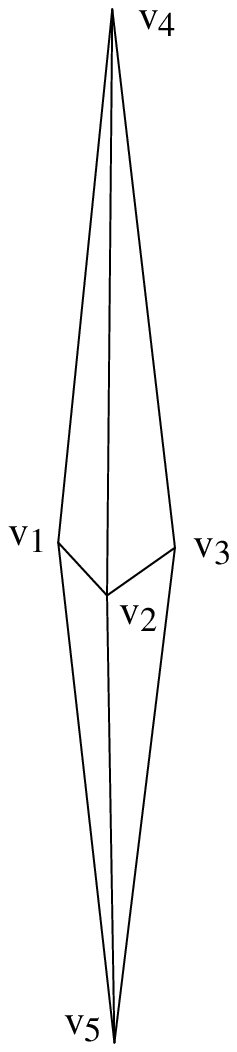}
\caption{A polytope $Q_2$ with two sharp vertices but no convex unfolding.}
\figlab{not.two.sharp}
\end{minipage}
\end{figure}

\begin{lemma}
There is a polytope with two sharp
vertices but with no convex unfolding.
\lemlab{not.two.sharp}
\end{lemma}
\begin{pf}
Our proof of this lemma is less straightforward, although the
example is simple.
Let $Q_2$ be the polytope formed by joining two copies of $Q_1$
from Lemma~\lemref{not.tetra} at their bases, as shown in
Fig.~\figref{not.two.sharp}. 
$Q_2$ is a $5$-vertex polytope, with vertices $v_1,\ldots,v_4$
as in $Q_1$, and $v_5$ the reflection of $v_4$ in the
central triangle $C = \triangle v_1 v_2 v_3$.
Again let $\e$ be the face angle incident to $v_4$ (and symmetrically $v_5$),
and choose $\e$ small so that only $v_4$ and $v_5$ are sharp vertices.

By Lemma~\lemref{cut.tree}(3), if $Q_2$ has a convex unfolding,
the cut tree must be a path with its two leaves at the two sharp
vertices.  By  Lemma~\lemref{cut.tree}(5), the path must be composed
of geodesics.  We now analyze the geodesics starting at $v_5$ and
show that there can be no piecewise simple geodesic path that
passes through all the vertices of $Q_2$.

We group the geodesics starting at $v_5$ into three classes:
\begin{enumerate}
\item The three geodesics that pass through a midpoint of an edge of
triangle
$C$.  Each of these passes through $v_4$
before encountering any of the other vertices, and so cannot serve
as the cut path.
\item The three geodesics that pass through a vertex of  $C$.
Because these vertices have low curvature ($2\e$), the geodesic must emerge
nearly headed toward $v_4$:  it cannot turn to hit another vertex of $C$
without creating a reflex angle in the unfolding.
If the geodesic goes directly to $v_4$, then again this cannot serve as
the cut path.  So it must head towards $v_4$ but miss it.  We
group this type of geodesic with the third class.
\item Geodesics that pass though an interior point of an edge
of $C$, but not the midpoint.
These geodesics all head toward $v_4$ but miss it.
\end{enumerate}
We now argue that all the geodesics in the third class (the only remaining
candidates) self-intersect after looping around $v_4$.
This will then establish the lemma.

\begin{figure}[htbp]
\centering
\includegraphics[width=11cm]{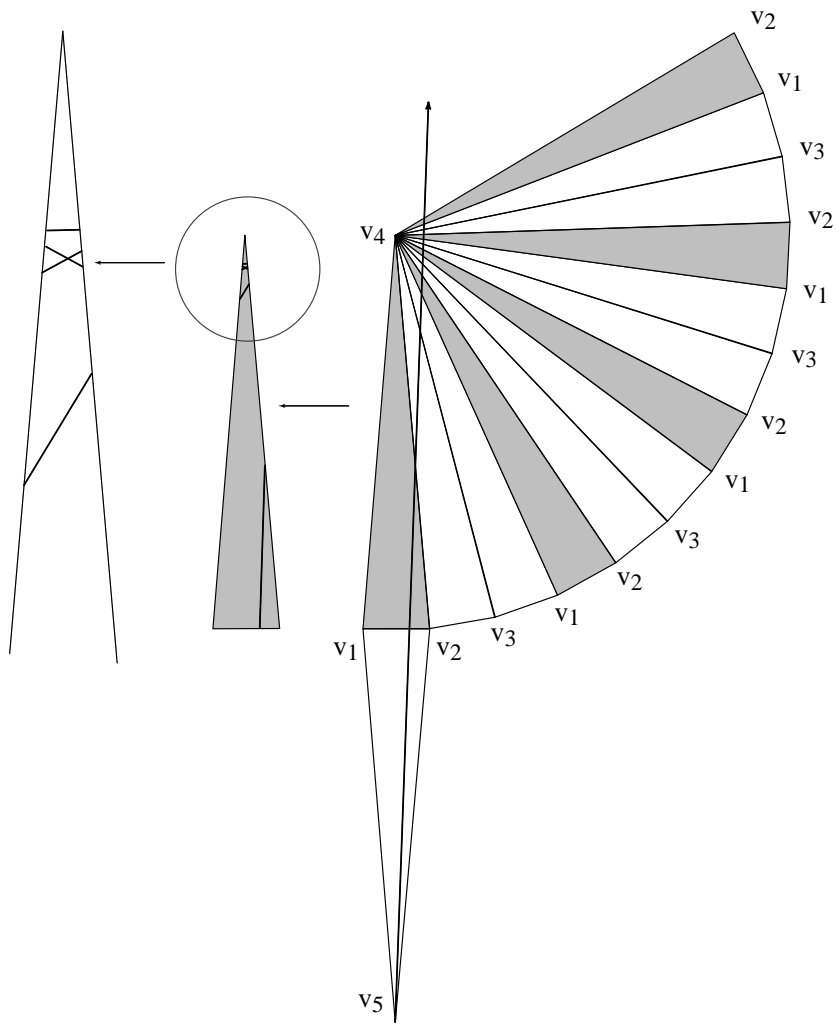}
\caption{A geodesic from $v_5$ that passes by $v_4$. The path of the 
geodesic on $\triangle v_1 v_2 v_4$ is shown to the left.}
\figlab{geodesic.crossing}
\end{figure}

An unfolding of a typical geodesic is shown in 
Fig.~\figref{geodesic.crossing}.
By choosing $\e$ small, we can arrange that every such geodesic crosses
several unfoldings of the three faces incident to $v_4$ before returning
back down to triangle $C$.  As can be seen from the copy of face
$\triangle v_1 v_2 v_4$ to the side, the path crosses each face several
times slanting one way, and then returns slanting the other way.
In the vicinity of the closest approach to $v_4$, the path
must self-cross.  We now establish this more formally.

\begin{figure}[htbp]
\centering
\includegraphics[width=0.7\linewidth]{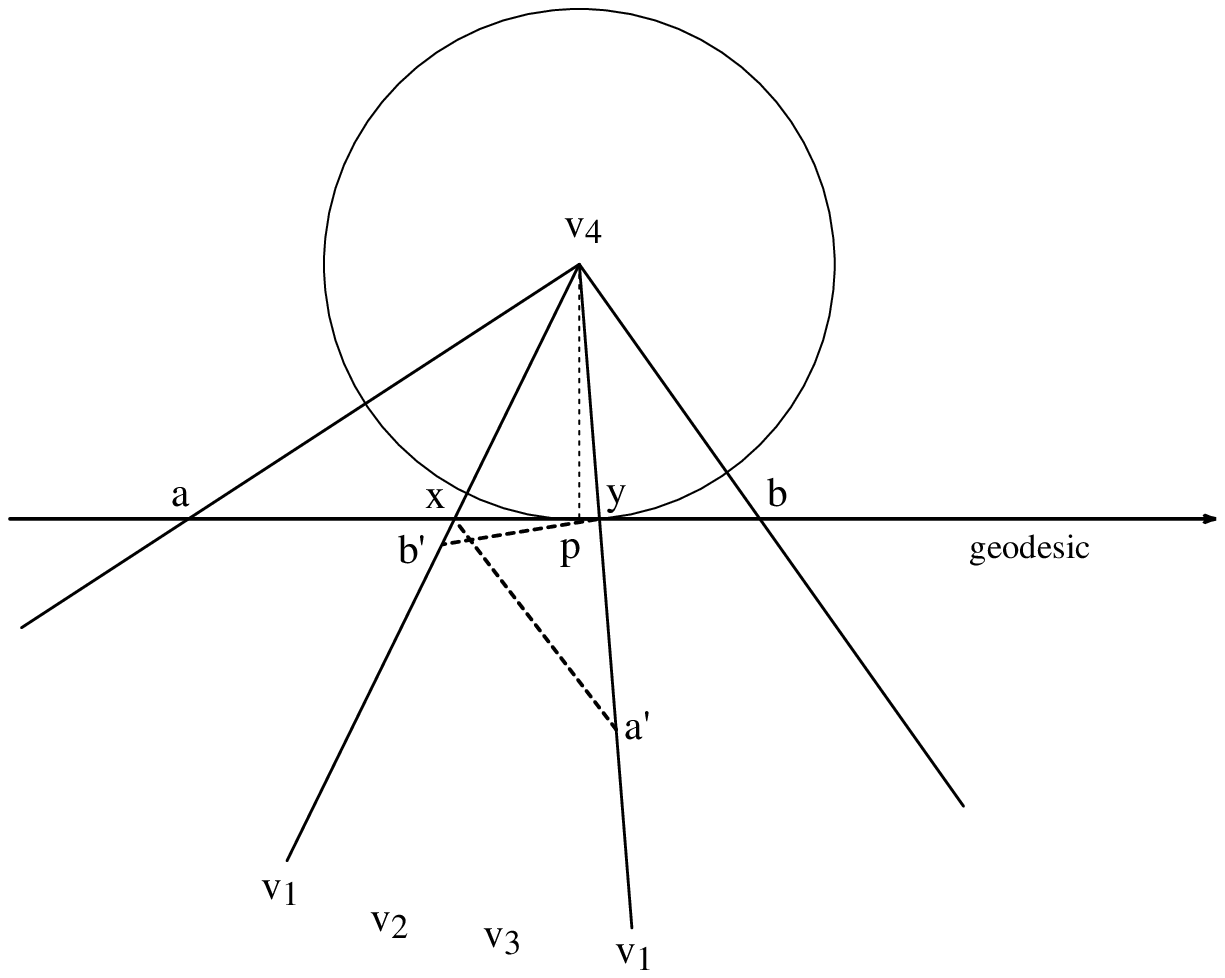}
\caption{$xa'$ and $yb'$ necessarily cross below the point $p$
of the geodesic closest to $v_4$.}
\figlab{X}
\end{figure}
Consider the unfolding of the three faces incident to $v_4$ (now viewed
as a unit) that includes the point $p$ of closest approach between the
geodesic and $v_4$; see Fig.~\figref{X}.
Let the geodesic cross the edge $v_1 v_4$ at points $a$, $x$, $y$, and $b$
in that order, with $xy$ including $p$.
Then $|v_4 b| > |v_4 x|$ and $|v_4 a| > |v_4 y|$,
because the distance from $v_4$ monotonically increases on either
side of $p$.
Thus the images $a'$ and $b'$ of $a$ and $b$ must fall below
$y$ and $x$ respectively in the figure.  Thus the geodesic must
cross somewhere in the stretch immediately before and after
the closest approach.

All we need for this argument to hold in general is for the geodesic to
cross three complete unfoldings of the three faces incident to $v_4$
before returning to the lower half of $Q_2$.  But this is easily
arranged by choosing $\e$ small.

We have shown that no geodesic starting from $v_5$ may serve as
a cut path for a convex unfolding.  Therefore $Q_2$ has no convex unfolding.
\end{pf}

\subsection{Necessary Conditions: Combinatorial Structure}

We now study the combinatorial structure of cut trees that lead to
convex unfoldings.
The following theorem is due to Shephard~\cite{s-cpcn-75},
although under different assumptions and with a different proof.%
\footnote{
        Shephard concludes that cut trees cannot have four leaves,
        an incorrect claim under our assumptions.
}

\begin{theorem}
If a polytope $Q$ of $n \neq 4$ vertices has a convex unfolding,
then the corresponding cut tree $T_C$ has two or three leaves:
it is either a path, or a `{\tt Y}' (a single degree-$3$ node).
If $n=4$, then additionally it may have four leaves, and
have the combinatorial structure of `{\tt +}' (a single degree-$4$ node),
or two degree-$3$ nodes connected by an edge, which we will call
a `{\tt I}'.
\theolab{cut.comb}
\end{theorem}
\begin{pf}
Let the cut tree $T_C$ unfold $Q$ to a convex polygon.
By Theorem~\theoref{two.sharp},
each leaf of $T_C$ must be at a sharp vertex $v$, and so have
curvature $\g(v) \ge \pi$.
If $T_C$ has more than four leaves $v$ (and therefore $n > 4$,
i.e., we are in the $n \neq 4$ case of the theorem claim),
$\sum_v \g(v) > 4 \pi$, which violates the Gauss-Bonnet theorem.
Therefore $T_C$ has no more than four leaves.
If $T_C$ has just two or three leaves, then the only possible 
combinatorial structures for $T_C$ are the two claimed in the theorem:
a path, and a `{\tt Y}'.
(Note that it is possible that $n=3$, when $Q$ is a doubly-covered
triangle.)

So assume that $T_C$ has exactly four leaves.
Because each leaf vertex is sharp,
$\sum_v \g(v) \ge 4 \pi$; on the other hand, we know the sum
over all vertices is equal $4 \pi$.  Therefore we know that
each leaf has curvature exactly $\pi$ and that
the leaves of $T_C$ are at the only vertices of $Q$.  Thus $n=4$
and $Q$ is a tetrahedron.
The only additional possible combinatorial structures for a tree with
four leaves are the two claimed in the theorem:
a `{\tt +}' and a `{\tt I}'.  Note that in both these cases, the internal
node(s) of $T_C$ are not at vertices of $Q$.
\end{pf}

A simple example of the `{\tt I}' possibility is shown in Fig.~\figref{rect.YY}.
If the rectangle is modified to become a square, the
`{\tt I}' becomes a `{\tt +}'.
\begin{figure}[htbp]
\centering
\includegraphics[width=0.5\textwidth]{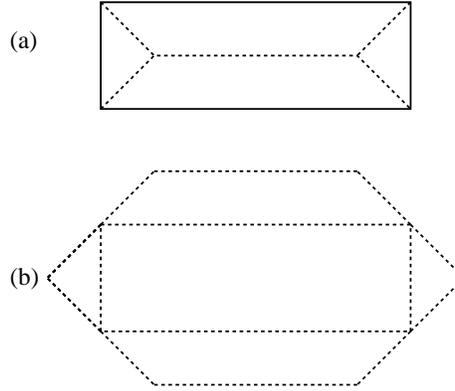}
\caption{A doubly-covered rectangle unfolds with a `{\tt I}' cut tree whose
leaves are the four rectangle corners.
Note all four rectangle corners are fold vertices.}
\figlab{rect.YY}
\end{figure}

\section{Counting Foldings: Gluing Trees}
\seclab{Counting.Gluings}
In this section we move beyond 
Lemma~\lemref{perim.halving}, which shows that every convex polygon
folds to a polytope, and explore how many different
ways there are to fold a given polygon, as measured by the
number of combinatorially distinct Aleksandrov gluing trees.
In Section~\secref{Noncongruent.Polytopes} we count instead the number of
distinct polytopes that might be produced from a given polygon.
In both cases, we will also examine the restriction
to convex polygons, which not surprisingly yields sharper results.

\subsection{Unfoldable Polygons}
We start with a natural and easily proved claim:

\begin{lemma}
Some polygons cannot be folded to any polytope.
\lemlab{unf}
\end{lemma}
\begin{pf}
Consider the polygon $P$ shown in Fig.~\figref{unf}.
\begin{figure}[htbp]
\centering
\includegraphics[height=3cm]{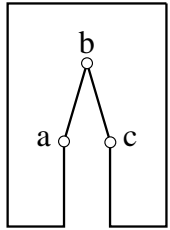}
\caption{An unfoldable polygon.}
\figlab{unf}
\end{figure}
$P$ has three consecutive reflex vertices $(a, b, c)$, with
the exterior angle $\b$ at $b$ small.  All other vertices
are convex, with interior angles strictly larger than $\b$.

Either the gluing ``zips'' at $b$, leaving $b$ a leaf of
$T_G$, or some other point(s) of $\bP$ glue to $b$.
The first possibility forces $a$ to glue to $c$, exceeding
$2\pi$ there; so this gluing is not Aleksandrov.
The second possibility cannot occur with $P$, because
no point of $\bP$ has small enough internal angle to fit at $b$.
Thus there is no Aleksandrov gluing of $P$.
\end{pf}

It is natural to wonder what the chances are that a random polygon
could fold to a polytope.
This is difficult to answer without a precise definition of
``random,'' but we feel any reasonable definition would
lead to the same answer:

\begin{conj}
The probability that a random polygon of $n$ vertices
can fold to a polytope approaches $0$ as $n \rightarrow \infty$.
\end{conj}
\begin{pf}
({\em Sketch\/}.)
Assume that random polygons on $n$ vertices satisfy two properties:
\begin{enumerate}
\item The distribution of the polygon angles approaches
the uniform distribution on the interval $(0,2\pi)$ as $n \rightarrow \infty$.
In particular, the number of reflex and convex vertices
approaches balance.
\item The distribution of polygon edge lengths approaches
some continuous density distribution.
\end{enumerate}
For large $n$, we expect $P$ to have $r=n/2$ reflex vertices.
Each of these reflex vertices $a$ faces one of two fates in the gluing
tree:  either it becomes a leaf by ``zipping'' at $a$;
or at least one convex vertex $b$ (of sufficiently small angle)
is glued to $a$.
The number of reflex vertices that can be zipped is limited
by Fact~\factref{4pi}:
if $a$ has angle $\a$, zipping there adds $2\pi-\a$ to the
curvature; but the total curvature is limited to $4\pi$.
Suppose we zip the largest $k$ angles out of the $r$ reflex vertices
(the largest angles increment the curvature the least).
Then one can compute that,
under the uniform angle distribution assumption,
these $k$ angles have an expected
curvature sum of
\begin{equation}
\frac{1}{2} \frac{\pi}{r} k^2 \; .
\end{equation}
(For example, for $r=100$, the largest $k=10$ have an expected curvature
sum of $\pi/2$.)
Limiting this to $4\pi$ implies that the expected maximum number
of reflex vertices that can be zipped without exceeding $4\pi$
curvature is
\begin{equation}
k \le 2 \sqrt{2 r} = 2 \sqrt{n} \; .
\end{equation}
(For example, for $r=1000$ reflex vertices, the largest $k=89$
lead to a curvature of $\approx 4\pi$.)
Thus, at most a small portion of the reflex vertices can be zipped;
the remainder (expected number: $n/2 - 2\sqrt{n}$) 
must be glued to convex vertices.  We now show that this gluing is
not in general possible.

Let $a$ be a reflex vertex with angle $\a$, 
and $b$ a convex vertex whose angle $\b$ satisfies $\b \le 2\pi-\a$,
so that $b$ can glue to $a$.
It could be that this gluing forces one or more reflex vertices
adjacent to $a$ or $b$ to glue to edges incident to $a$ or $b$,
in which case the gluing is not possible (i.e., it is not an
Aleksandrov gluing).
If the adjacent vertices are convex, and/or the edge lengths are
such that the gluing is Aleksandrov, then, in general, two new reflex vertices
are created, as is illustrated in
Fig.~\figref{random}.
\begin{figure}[htbp]
\centering
\includegraphics[width=0.75\linewidth]{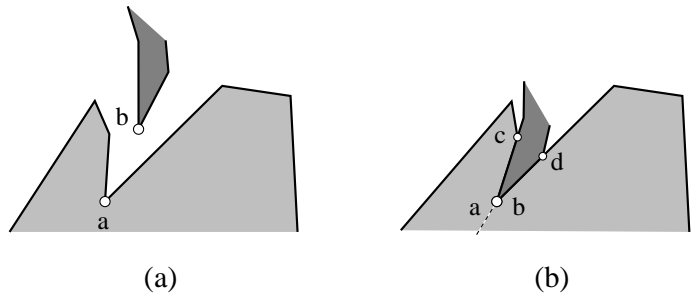}
\caption{(a) Reflex vertex $a$, convex vertex $b$; 
(b) Two new reflex vertices produced by gluing $b$ to $a$.}
\figlab{random}
\end{figure}

To be more precise, let $A = |a a_1|$ be the length of the edge
incident to $a$ which is glued to the length $B = |b b_1|$ of
an edge incident to $b$.
If $A > B$ and $b_1$ is reflex, the gluing is {\em not\/} Aleksandrov;
but if $b_1$ is convex, a new reflex vertex is created at $b_1$.
Symmetrically, if $B > A$ and $a_1$ is reflex, the gluing is
not possible; but if $a_1$ is convex, a new reflex vertex is
created at $a_1$.  The only circumstance in which the gluing
is Aleksandrov and a new reflex vertex is not created is when
$A = B$ and both $a_1$ and $b_1$ are convex with an angle sum
of no more than $\pi$.

Under the assumption that the edge lengths approach some continuous
distribution, the probability that two lengths match exactly
approaches $0$.  Thus we conclude that gluing convex vertices
to reflex vertices does not remove reflex vertices, but rather
creates new ones in shorter polygonal chains, one new reflex
vertex in each of the two chains produced by the gluing.
Note that gluing several convex vertices to one reflex vertex
does not change matters: we can view the first convex vertex
as simply leaving a reflex remainder, and argue as above.

Thus, any gluing of a random polygon for large $n$
will lead to shorter and and shorter chains ``pinched'' between reflex-convex
gluings, each of which will contain at least one reflex vertex
(actually, two reflex vertices for those pinched on both sides).
Eventually these chains reach the point where either there are
no convex vertices that fit into the reflex vertex gap, or
there are no convex vertices at all.  In either case, the chain
cannot be glued:  the reflex vertex would have to glue to a point
interior to an edge, violating the Aleksandrov condition that
no point have more than $2\pi$ glued angle.
\end{pf}

The proof above hinges on the unlikeliness of matching
edge lengths.  It is therefore natural to wonder if
the same result holds for polygons all of whose edge lengths
are the same.  Again we believe it does:

\begin{conj}
The probability that a random polygon of $n$ vertices,
all of whose edges have unit length,
can fold to a polytope, approaches $0$ as $n \rightarrow \infty$.
\end{conj}
\begin{pf}
({\em Sketch\/}.)
Assume a model of random polygons such that the angles are
probabilistically independent and uniformly distributed
in $(0,2\pi)$ as $n \rightarrow \infty$.
The restriction to unit edge lengths means that all gluings are
vertex to vertex (no vertex is ever glued to the interior of an edge).
The gluing is Aleksandrov iff the angles glued together sum to at most
$2\pi$
everywhere.

Consider gluing two vertices to one another.  Because their angles
are independent, the chance that the gluing is legal is $1/2$ (the sum of
their distributions is uniform between $0$ and $4\pi$). Gluing $k$ pairs then has
a $1/2^k$ chance of being Aleksandrov.

As in the above proof sketch, the gluing tree cannot have too
many leaves.  Zipping just $2 \sqrt{n}$
reflex vertices uses up all $4 \pi$ of
curvature. 
So the number of leaves is only about $2 \sqrt{n}$. As we will see
in Theorem~\theoref{gluing.upper} below,
specifying a ``source'' for each leaf pins down the whole
tree structure.  So by selecting $4 \sqrt{n}$ vertices for the leaves and
their sources, the gluing tree is determined.

Therefore we should compare the number of different gluing
trees,
\begin{equation}
\left(
	\begin{array}{c}
		n \\ 
		4 \sqrt{n}
	\end{array}
\right)
\eqlab{num}
\end{equation}
to the probability that each one is Aleksandrov,
\begin{equation}
\frac{1}{2^{n-4 \sqrt{n}}}
\eqlab{den}
\end{equation}
Note here we conservatively only concern ourselves with
degree-$2$ vertex-to-vertex gluings; junctions of degree $d > 2$ have
a lower probability of summing to no more than $2\pi$.
We also ignore the change to the angle distribution 
caused by the removal of the leaf vertices.

Using Stirling's approximation shows that the 
$\log$ of Eq.~\eqref{num} grows as
$2 \sqrt{n} \log n$; but the $\log$ of
Eq.~\eqref{den} grows as $n$. So their ratio
approaches $0$ as $n \rightarrow \infty$.
\end{pf}

We leave these results on random polygons as conjectures,
as it would require a more precise definition of what
constitutes a random polygon, and more careful probabilistic
analyses, to establish them formally. 

\subsection{Lower Bound: Exponential Number of Gluing Trees}
In contrast to the likely paucity of foldable polygons, some
polygons generate many foldings.

\begin{theorem}
For any even $n$, there is a polygon $P$ of $n$ vertices
that has $2^{\Omega(n)}$ combinatorially distinct
Aleksandrov gluings.
\theolab{star}
\end{theorem}
\begin{pf}
The polygon $P$ is illustrated in Fig.~\figref{star}(a).
\begin{figure}[htbp]
\centering
\includegraphics[width=\linewidth]{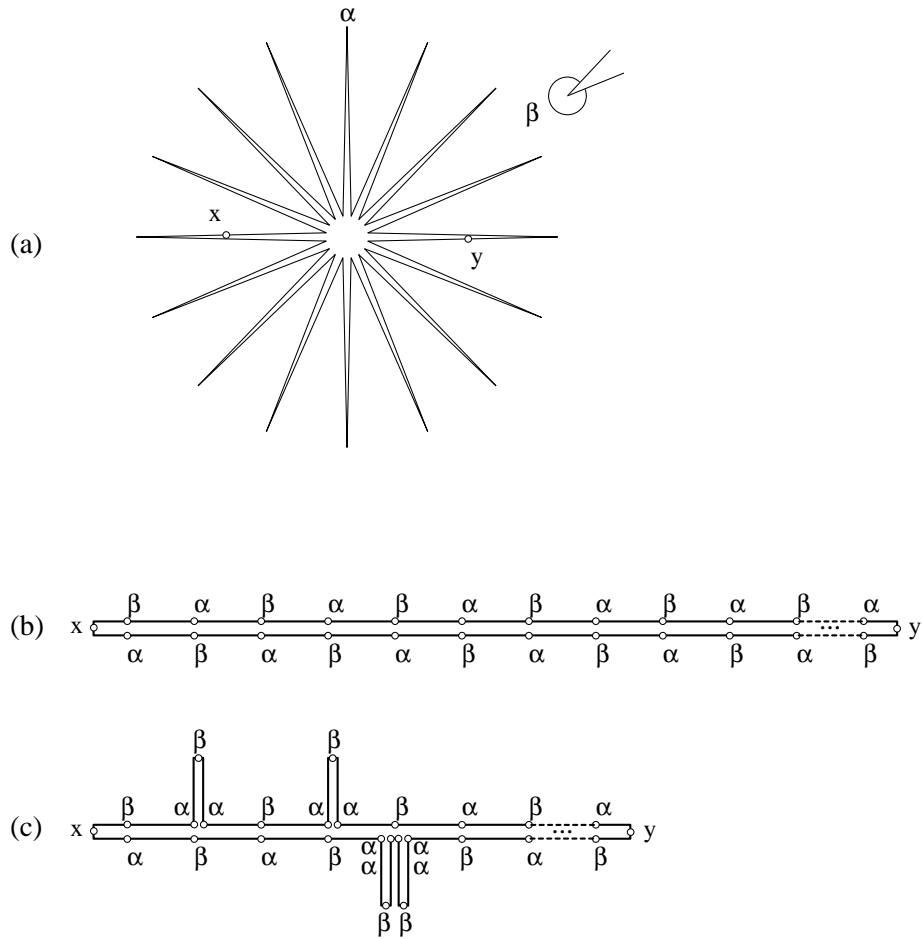}
\caption{(a) Star polygon $P$, $m=16$, $n'=32$, $n=34$.
(b) Base gluing tree.
(c) A gluing tree after several contractions.}
\figlab{star}
\end{figure}
It is a centrally symmetric star, with $m$ vertices, $m$ even,
with a small convex angle $\a \approx 0$, 
alternating with $m$ vertices
with large reflex angle $\b < 2\pi$.
All edges have the same (say, unit) length.
We call this an $m$-star.
We first specify the constraints on $\a$ and $\b$.

$P$ has $n'=2m$ vertices (ignoring $x$ and $y$, to be described shortly).
So $m(\a+\b) = (n'-2)\pi$, which implies that 
\begin{equation}
\a + \b = (1 - \frac{1}{m}) 2 \pi \;. \eqlab{a+b}
\end{equation}
We choose $\a$ small enough so that $m$ copies of $\a$ can
join with one of $\b$ and still be less than $2\pi$:

\begin{equation}
m \a + \b < 2 \pi  \;. \eqlab{ma}
\end{equation}
Substituting this relationship into Eq.~\eqref{a+b} and solving
for $\a$ yields:
\begin{equation}
\a < \frac{2 \pi}{ m(m-1) } \;. \eqlab{a}
\end{equation}

Now we add two vertices $x$ and $y$ at the midpoints of edges,
symmetrically placed so that $y$ is half the perimeter around
$\bP$ from $x$.  Let $n=n'+2$ be the total number of vertices of $P$.

The ``base'' gluing tree is illustrated in 
Fig.~\figref{star}(b).
$x$ and $y$ are fold vertices of the gluing.
Otherwise, each $\a$ is matched with a $\b$.
Because all edge lengths are the same, and because 
$\a+\b < 2\pi$ by Eq.~\eqref{a+b},
this path is an Aleksandrov gluing.
We label it $T_{00\cdots0,00\cdots0}$,
where $m/2$ zeros $00\cdots0$ represent the top chain, and another $m/2$
zeros represent the bottom chain.

The other gluing trees are obtained via ``contractions''
of the base tree.
A {\em contraction\/} makes any particular $\b$-vertex 
not adjacent to $x$ or $y$
a leaf of the tree by gluing its two adjacent $\a$-vertices
together.  Label a $\b$-vertex $0$ or $1$ depending
on whether it is uncontracted or contracted
respectively.  Then a series of contractions can be identified
with a binary string.
For example, Fig.~\figref{star}(c) displays the
tree $T_{010100\cdots,00110\cdots0}$.
Note that $k$ adjacent contractions result in $2k$ 
$\a$-vertices glued together.

We now claim that if the number of contractions in the top chain
is the same as the number in the bottom chain
(call such a series of contractions {\em balanced\/}),
the resulting
tree represents an Aleksandrov gluing.
Fix the position of $x$ to the left,
and contract leftwards, as in  Fig.~\figref{star}(c).
Then it is evident that the alternating ``parity'' pattern of
$\a$'s and $\b$'s is not changed by contractions.
Ignoring the arcs attached to the central path,
each contraction replaces $\a \rightarrow 2\a$,
and shortens the path by $2$ units.
Because the contraction shortens by an even number of units,
it does not affect the parity pattern.
If the top and bottom chains are contracted the same number of
times (twice each in (c) of the figure), then their lengths
are the same.

Thus after a balanced series of contractions, we have a number
of $\b$-leaves, and gluings of $2k$ $\a$-vertices to one $\b$-vertex.
The $\b$-leaves are legal gluings because $\b < 2\pi$.
Because there are $m/2-1$ contractible $\b$-vertices in each chain, the longest
series of adjacent contractions is $m/2-1$.  So $k \le m/2 - 1$, and
$2k < m$.  Eq.~\eqref{ma} then shows that each gluing produces
less than $2\pi$ angle, and so is Aleksandrov.

Finally, we bound the number of gluings.
There are $2^{m/2-1}$ binary numbers of $m/2-1$ bits.
Thus there are this many ways to contract the top chain.
The bottom chain must be contracted with the same number
of $1$'s for a balanced series.  Rather than count this
explicitly, we simply note that $P$ has
at least  $2^{m/2-1}$ 
Aleksandrov gluings, and
because $P$ has $n=n'+2=2m+2$ vertices,
$\Omega(2^{m/2-1}) = \Omega(2^{(n-6)/4}) = 2^{\Omega(n)}$.
\end{pf}

\begin{figure}[htbp]
\centering
\includegraphics[width=0.75\linewidth,angle=-90]{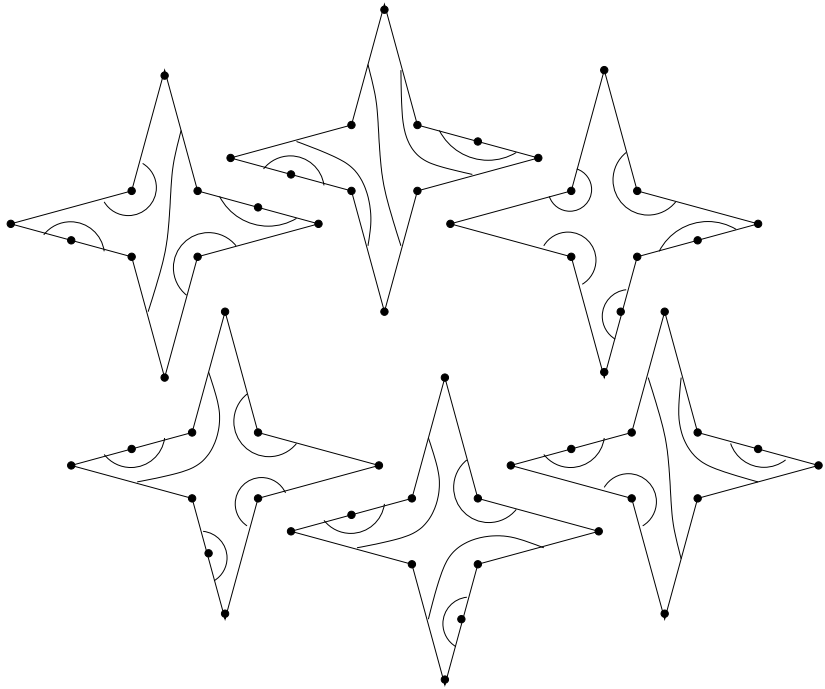}
\caption{Six gluing patterns for a $4$-star.}
\figlab{4star_gluings}
\end{figure}
Fig.~\figref{4star_gluings} shows six gluings
of a $4$-star.
The first two in the top row correspond to the
perimeter-halving construction used in the proof.
By Aleksandrov's theorem, each corresponds to a
unique polytope, but
as mentioned in Section~\secref{Aleksandrov},
we do not know how to compute the 3D structure
of these polytopes.
Nevertheless, our hand-exploration suggest that all
fold to noncongruent polytopes, each with the combinatorial
structure of the regular octahedron.
Two of our conjectured crease patterns are shown in 
Fig.~\figref{crease}.

\begin{figure}[htbp]
\centering
\centering
\includegraphics[width=0.45\linewidth]{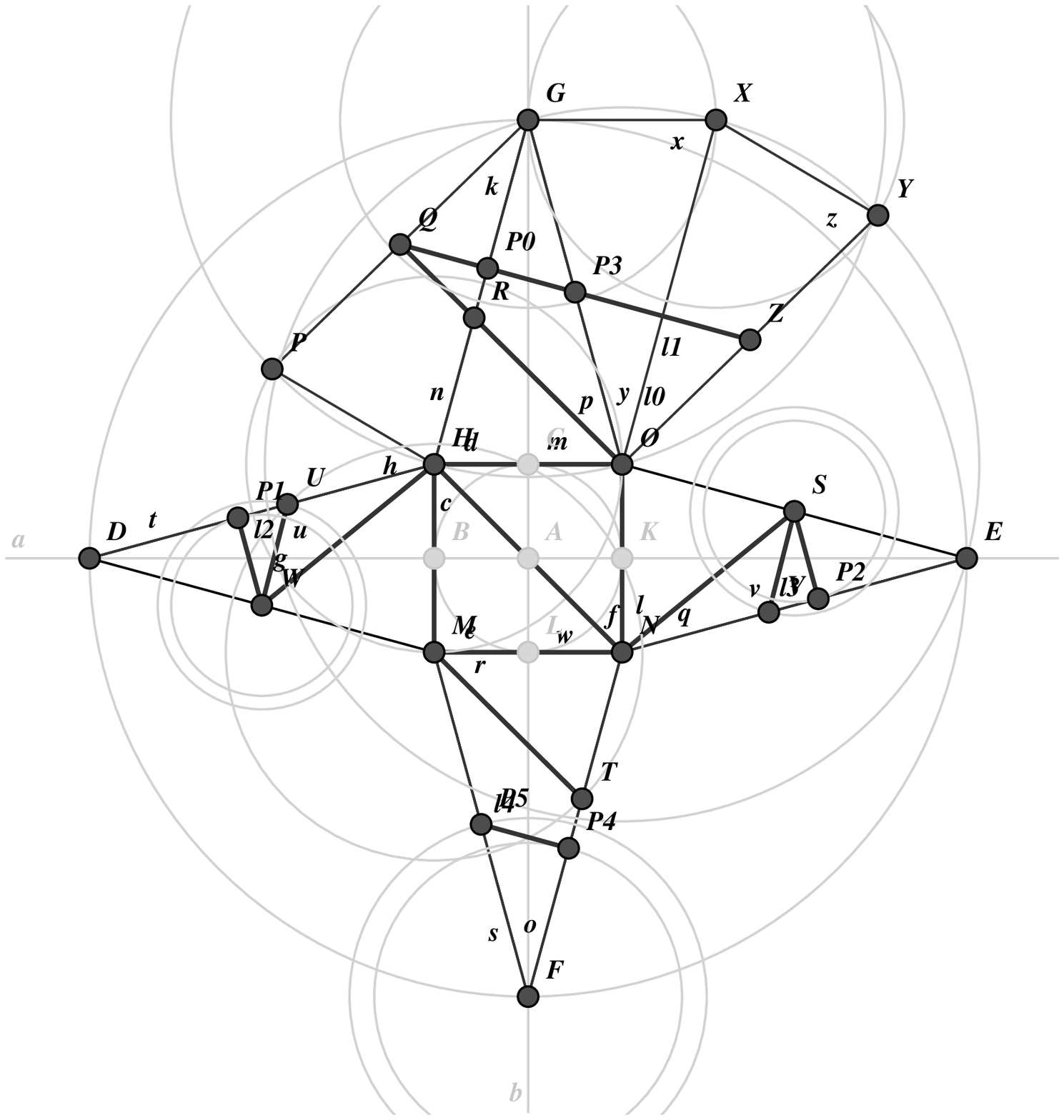}
\hspace{5mm}%
\includegraphics[width=0.45\linewidth]{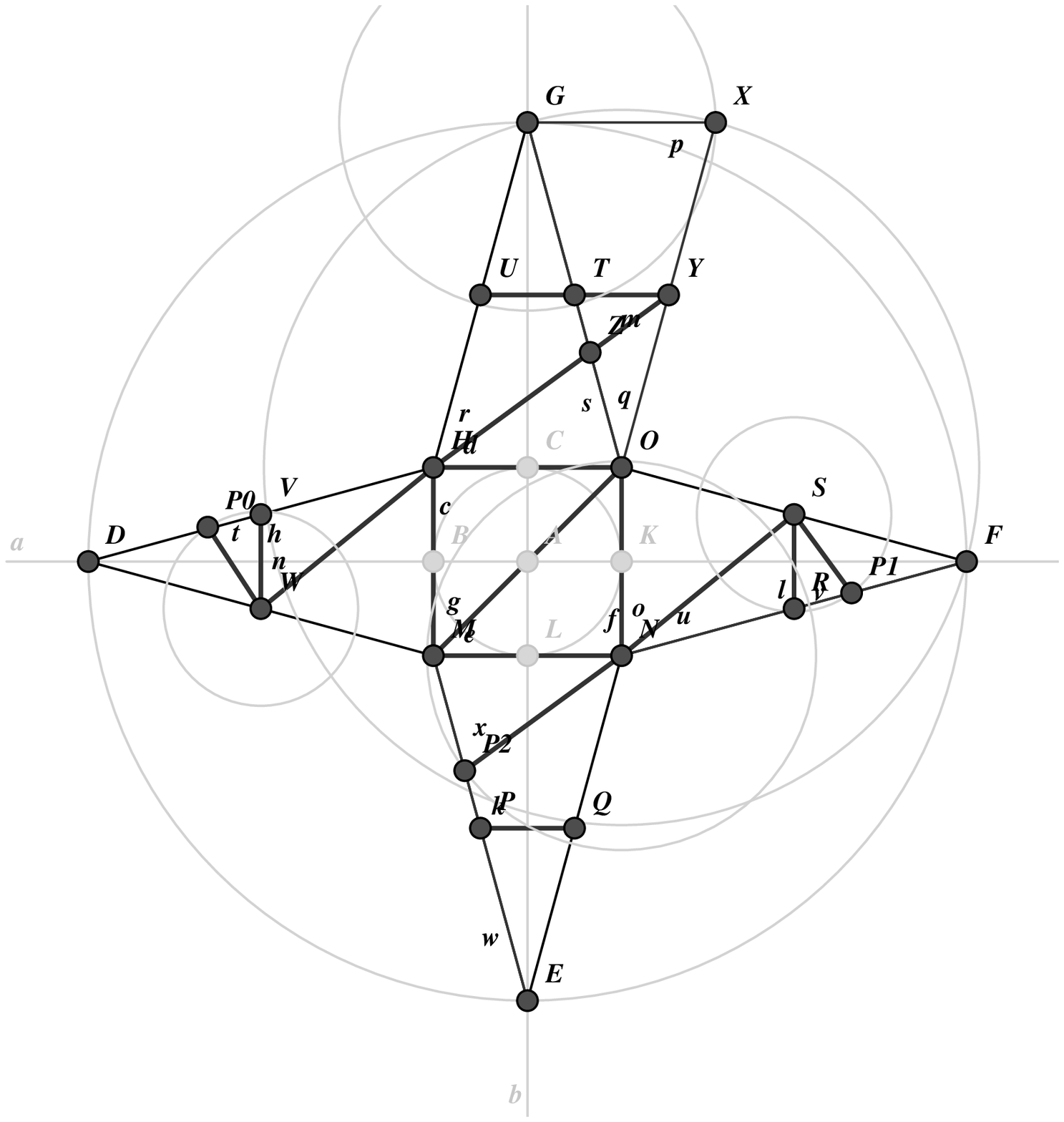}
\caption{Conjectured crease patterns for the first
two gluing patterns in the top row of
Fig.~\figref{4star_gluings}.
[Constructions performed in Cinderella.]}
\figlab{crease}
\end{figure}

\subsection{Upper Bound: Few Leaves}
Our goal is now to provide upper bounds on 
the number of gluings, both for arbitary polygons and for
convex polygons.  Both will rely on upper bounds for
gluing trees with a small number of leaves.
Let a gluing tree $T_G$ have $\l$ leaves.
In this section, we prove results for $\l=2$ and $\l=3$.
We then use these to obtain a general upper bound
in Section~\secref{Gluing.Upper.General}, and a bound for convex polygons
in Section~\secref{Gluing.Upper.Convex}.
In between, we summarize the structural properties of
gluing trees in Section~\secref{Gluing.char}.

It will sometimes be easier to work with ``gluing instructions''
rather than with gluing trees.
Toward that end, we define
the combinatorial
type of a gluing.
Again let polgyon $P$ have vertices $v_i$ and edges $e_i$,
labeled counterclockwise,
The {\em combinatorial type\/} 
$\G_G$ of a gluing $G$ specifies
to which vertex or edge of $P$ each vertex of $P$ glues
via a set of ordered pairs:
$\G_G = \{ (v_i, z_j) \}$, where $z_j$ is either $v_j$ or $e_j$,
the first element $j$ to which $v_i$ glues counterclockwise
around $\bP$.  If $v_i$ is a leaf of the cut tree, then
the pair $(v_i,v_i)$ is included; otherwise $v_i$ must glue to
an element different from itself.
For example, 
the combinatorial type of the gluing illustrated earlier in
Fig.~\figref{tri.tetra}a is
$$
\{
(v_1,v_3),
(v_2,v_2),
(v_3,e_3)
\}
$$

We now prove that the combinatorial type of a gluing
determines the gluing tree.

\begin{lemma}
The combinatorial type $\G_G$ of a gluing 
$G$ determines the gluing tree $T_G$.
\lemlab{gluing.tree}
\end{lemma}
\begin{pf}
A node of degree~$2$ of $T_G$ is directly labeled in $\G_G$
as either $(v_i,v_j)$ or $(v_i,e_j)$.
It is only nodes of degree~$\neq 2$ for which $T_G$ contains information
not immediately supplied by $\G_G$.
Nodes of degree~$1$ (leaves) of $T_G$ 
correspond to two possible types of gluings:
either $(v_i,v_i)$, which are directly labeled in $\G_G$, or
fold vertices, a vertex produced by folding 
at a point $x$ in the interior of an edge $e_j$.
(Cf.~Fig.~\figref{perim.halving} for an example of fold vertices.)
Fold vertices can be identified in $\G_G$ as gluings of $v_i$ to
either $e_i$ or $e_{i-1}$: gluing to an incident edge necessarily
implies a fold vertex on that edge.
Or $v_i$ can be glued
to the next vertex, folding the edge in half.
In Fig.~\figref{tri.tetra}, the pair $(v_3,e_3)$ identifies
fold vertex $x$ as labeled with $e_3$;
that $v_1$ also glues to incident edge $e_3$ is known after the
degree~$3$ node's labels are determined.

Nodes of degree~$d>2$ in $T_G$ have $d$ labels.
Because every such node can involve at most one edge
(because two edges glued to a point already gives an angle of $2\pi$
there, and the other
elements glued to the same point would cause the angle sum to exceed this),
the labels can be gathered by following the gluings counterclockwise:
$$(v_{i_1},v_{i_2}), (v_{i_2},v_{i_3}), \ldots, 
(v_{i_{d{-}2}},v_{i_{d{-}1}}), (v_{i_{d{-}1}},e_j) \,.$$
In Fig.~\figref{tri.tetra}, the node at point $z$ has labels
$\{v_1,v_3,e_3\}$, which can be identified from
the pairs
$(v_1,v_3),(v_3,e_3)$ of $\G_G$.
\end{pf}

\noindent
This lemma permits us to count gluing trees by counting combinatorial
types of gluings.

\begin{lemma}
A polygon $P$ of $n$ vertices has
$\Theta(n^2)$ different gluing trees of two leaves,
i.e., paths.
\lemlab{count.path}
\end{lemma}
\begin{pf}
View $\bP$ as rolling continuously between the two leaves $x$ and $y$,
like a conveyor belt or tank tread.
Each specific position corresponds to a perimeter-halving
gluing $G$ (Fig.~\figref{perim.halving}).
The combinatorial type $\G_G$ changes each time a vertex $v_i$
either passes another vertex $v_j$, or becomes the leaf $x$ or $y$.
Each such event corresponds to two distinct types: the type at the
event, and the type just beyond it:  e.g., $(v_i,v_j)$ and $(v_i,e_j)$.
So counting events undercounts by half.
If we count the possible pairs $(v_i,v_j)$ 
for all $i \neq j$, 
we will double count each type:
the event $(v_i,v_j)$ leads to the same type as
$(v_j,v_i)$.
The undercount by half and overcount by double cancel;
thus $n(n-1)$ is the number of types without a vertex at a leaf.
Adding in the $n$ possible $(v_i,v_i)$ events, each of which
leads to two types, yields
an upper bound of $n(n-1) + 2n = O(n^2)$ on the number of
combinatorial types.

A lower bound of $\Omega(n^2)$ is achieved by the
example illustrated in 
Fig.~\figref{lower.bounds}(a).
Here $n/2$ vertices of $P$ are closely spaced within a
length $L$ of $\bP$, and $n/2$ vertices are spread out by more
than $L$ between each adjacent pair. Then each of the latter
vertices (on the lower belt in the figure) can be placed between
each pair of the former vertices (on the upper belt), yielding
$n^2/4$ distinct types.  This example can be realized geometrically
by making the internal angle at each vertex nearly $\pi$,
i.e., by a convex polyon that approximates a circle.

Lemma~\lemref{gluing.tree} shows that the bound just
obtained of $\Theta(n^2)$ on the number of combinatorial types
applies as well to the number of gluing trees.
\end{pf}

\begin{figure}[htbp]
\centering
\includegraphics[width=8cm]{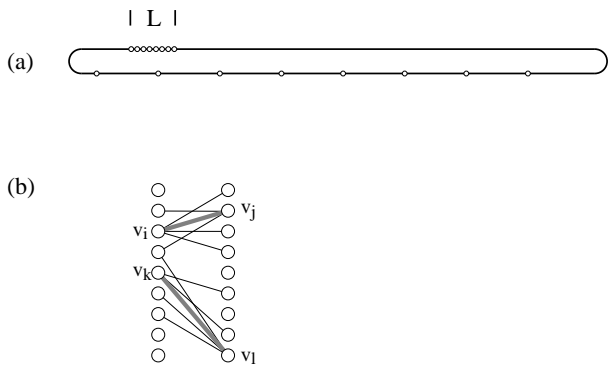}
\caption{(a) $\Omega(n^2)$ combinatorial types can be achieved
by rolling this perimeter ``belt'';
(b) There are only $O(n)$ possible disjoint vv-pairings.}
\figlab{lower.bounds}
\end{figure}

\begin{lemma}
A polygon $P$ of $n$ vertices folds to at most
$O(n^4)$ different gluing trees of three leaves,
i.e., `{\tt Y}'s.
\lemlab{count.Y}
\end{lemma}
\begin{pf}
Observe that the degree-$3$ node of the `{\tt Y}' is either
comprised by the gluing two vertices and an edge together
(call this {\em type-vve}),
or three vertices ({\em type-vvv}).
It is not possible to glue two or more edges together without
violating the $\le 2\pi$ angle restriction of an Aleksandrov
gluing.

There are $O(n^3)$ possible type-vvv nodes for the `{\tt Y}'.
Once this type of node is specified, the entire gluing tree is determined,
so this bounds the number of `{\tt Y}'s with type-vvv nodes.
Now consider type-vve nodes.
There are $O(n^2)$ possible vv-gluings, which determine
one branch of the `{\tt Y}'.  The remainder of the `Y' can be
viewed as a path between its leaves; essentially this view
corresponds to a conveyor belt with an appendage.
Applying Lemma~\lemref{count.path} yields a bound of $O(n^4)$.

\end{pf}

We leave open the question of whether this bound is tight.
We will improve it for convex polygons in Section~\secref{Gluing.Upper.Convex}.

\subsubsection{Four Fold-Point Gluing Trees}
We now embark on a study of a special case that will play
two roles: in the proof of our main combinatorial
upper bound, Theorem~\theoref{gluing.upper},
and in counting noncongruent polytopes in Section~\secref{Noncongruent.Polytopes}.
Define a {\em four fold-point gluing tree\/} to be a gluing
tree with (at least) four leaves, each fold points, i.e., creases in the
interior of polygon edges leading to polytope vertices of
curvature $\pi$.
We have already encountered one such tree in Fig.~\figref{zed.tetra2}(b).
We start with this straightforward lemma.

\begin{lemma}
A four fold-point gluing tree must have exactly four leaves,
and so have
combinatorial structure
`{\tt +}' or `{\tt I}'.
\lemlab{four.4}
\end{lemma}
\begin{pf}
Because each fold point leads to a
vertex of the resulting polytope $Q$ which
has curvature $\pi$, Fact~\factref{4pi}
implies that all the curvature of the polytope is at the
four fold vertices.
Thus all vertices of $P$ must glue to points that have
total angle $2\pi$, so that the curvature there is zero.

A leaf of a gluing tree cannot have zero curvature.
This is because a leaf is either a fold point (curvature $\pi$)
or a ``zipped'' polygon vertex $v$.  The only way to achieve zero
curvature at a zipped vertex is to have an internal polygon
angle at $v$ of $2\pi$.  But this violates simplicity of $P$:
all internal angles are strictly less than $2\pi$.

Therefore, a four-fold gluing tree must have exactly four leaves.
So there are only two possible combinatorial structures:
`{\tt +}' and `{\tt I}' (as in Lemma~\lemref{cut.tree}).
\end{pf}

Before counting the number of gluing trees, we detail one example
that will be the basis for the remainder of our analysis.
Start with an $L \times W$ rectangle $P$, and fold it as follows.
Glue the two opposite edges of length $W$ together to form a
cylinder.  Now glue the bottom rim of the cylinder to itself
by creasing at two diametrically opposed points $x_1$ and $y_1$.  
Similarly
glue the top rim to itself by creasing at two points $x_2$ and $y_2$.
The gluing tree is of structure `{\tt I}': see
Fig.~\figref{I}.
\begin{figure}[htbp]
\centering
\includegraphics[width=0.3\linewidth]{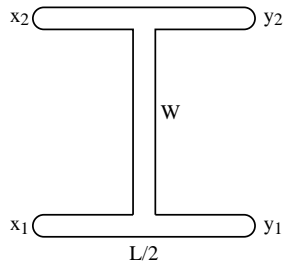}
\caption{`{\tt I}' gluing tree for an $L \times W$ rectangle.}
\figlab{I}
\end{figure}
It is easy to see this is an Aleksandrov gluing.  Note both internal
nodes of the gluing tree glue two $\pi/2$ rectangle corners
to the interior of an $L$-edge; so the angle sum there is $2\pi$.
The gluing is Aleksandrov
even if the crease
points on the top and bottom are not located at corresponding
points on their rims.  In particular, identify the points $x_i$ with
their distance from the rectangle corner to the left.  If $x_1 = x_2$,
then the crease points correspond, and the gluing produces a
flat, $L/2 \times W$ rectangle.  If $x_1 \neq x_2$, the gluing is
still Aleksandrov, but the ``twist'' in the gluing results in
a nondegenerate tetrahedron, with all vertices of curvature $\pi$.
Let $x = |x_2 - x_1|$ characterize amount of the twist,
with $x=0$ representing no twist.  

Because the $1$-skeleton
of a tetrahedron is combinatorially $K_4$, each vertex is
adjacent to all the others via polytope edges.
This makes it trivial to decide the structure of the polytope
$Q_x$ created by this rectangle gluing with twist $x$.
The six distances between pairs of vertices are easily computed
from the gluing,
and each represents an edge length.
These six lengths uniquely determine the 3D shape of the tetrahedron.
It is not difficult to compute 3D vertex coordinates from the
six lengths, and
we have written code for this computation.
An example is shown in
Fig.~\figref{tetra}.
Here a $2 \times 2$ rectangle is folded with a variety of
different twists $x$.
For both $x=0$ and $x=1$, the result is a flat $1 \times 2$
rectangle, with a smooth interpolation between for $0 < x < 1$.
\begin{figure}[htbp]
\centering
\includegraphics[width=\linewidth]{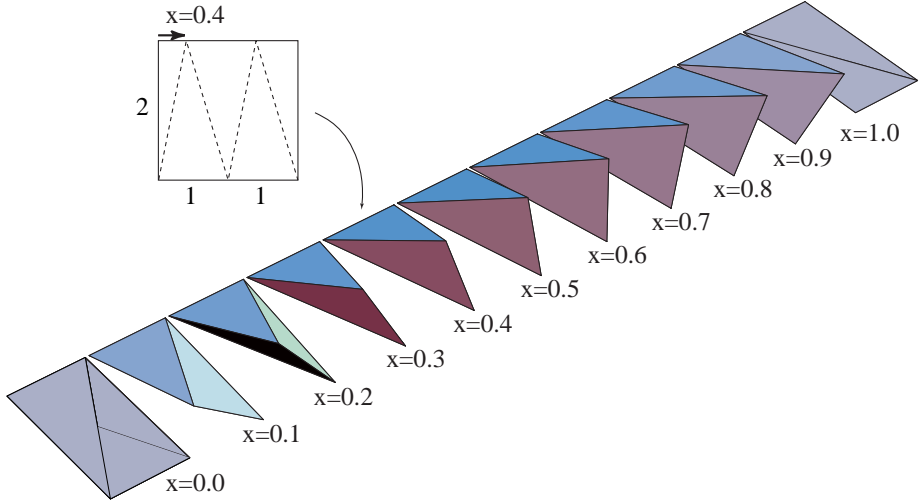}
\caption{Tetrahedra formed by folding a rectangle according
to the gluing tree shown in Fig.~\figref{I}.}
\figlab{tetra}
\end{figure}

We have proven this lemma:
\begin{lemma}
Any rectangle may fold via a `{\tt I}' gluing tree
to a uncountably infinite number of noncongruent tetrahedra.
\lemlab{tetra}
\end{lemma}
\begin{pf}
Two tetrahedra with different edges lengths are not congruent.
The edge lengths of $Q_x$ for twist $x$
are $L/2$ (twice), 
$u(x) = \sqrt{x^2 + W^2}$ (twice),
and $v(x) = \sqrt{(1-x)^2 + W^2}$ (twice).
For $a \neq b$, $u(a) \neq u(b)$;
and for $x < L/4$, $u(x) \neq v(x)$.
Thus the number of noncongruent tetrahedra is at least the
number of distinct $x \in [0, L/4)$, which is
nondenumerable.
\end{pf}

We return now to the task of upper-bounding the number of
four fold-point gluing trees possible for a polygon of $n$ vertices.
Although we do not at this point have tight bounds, they
suffice for our purposes in the next section.

Define a {\em conveyor belt\/} (or just {\em belt\/})~\label{rolling.belt}
in a gluing tree to be a path between
two leaf fold points.
Let a belt have fold points $x$ and $y$, with $x$
an interior point of edge $e$.
A belt can {\em roll\/} if there is a nonzero-length
interval $I \subset e$ such that for every
$x \in I$, the belt folded at $x$ is an Aleksandrov
gluing.
A belt could instead have a finite number of distinct
gluings, perhaps just one.
We first show that rolling belts must be vertex-free in
four fold-point trees.

\begin{lemma}
A rolling belt in a four fold-point tree $T$ cannot contain any
vertices except those at the attachment points to other
branches of $T$.
\lemlab{four.belt}
\end{lemma}
\begin{pf}
Suppose to the contrary that a rolling belt
contains at least one vertex $v$ in its interior,
i.e., not at an attachment point.
Because under our definition, all vertices of $P$ are essential,
the internal angle at $v$ is different from $\pi$.
Let $x \in I$ be a particular fold point that determines the
gluing of the belt.
In this position, $v$ must match up with another vertex $v'$ with
supplementary angle.
Rolling the belt in a neighborhood of $x$ breaks the match, leaving
both $v$ and $v'$ glued to points internal to an edge.
At these points, the curvature is greater than zero,
violating the fact that all curvature at a four fold-point gluing
are concentrated at the leaves.
\end{pf}

Note that the angles at the attachment points must be $\pi$.

\begin{lemma}
A belt in a four-fold gluing tree $T$ has at most $O(n)$
combinatorially distinct gluings.
\lemlab{four.n}
\end{lemma}
\begin{pf}
Let $B$ be a belt with attachment points $a$ and $b$.
Note that because each attachment point is an internal node
of $T$, the limited structural possibilities established
in Lemma~\lemref{four.4} allow only one or two attachment points.
Consider two cases:
\begin{enumerate}
\item $B$ can roll.  Then by Lemma~\lemref{four.belt},
$B$ contains no internal vertices.  Thus its only vertices are
at $a$ and $b$. Rolling can produce just two combinatorially distinct
positions of the belt.
\item $B$ cannot roll.  Then $B$ can assume a finite number of
possible positions.  
Define a {\em kink\/} in $B$ to be either a vertex, or an
attachment point at which the angle is different from $\pi$,
composed of two glued vertices.
The kinks must match up in pairs.
Matching one pair forces the remaining matches.  Thus
This can be seen by distributing the kinks around a topological
circle representing $B$.  Once one chord $v_1 v_i$ is drawn in
this circle, all other chords are forced by the pairwise matching
requirement.
Because there are only $m-1 < n$ choices for the mate for $v_1$,
$B$ has only $O(n)$ legal gluings.
\end{enumerate}
\end{pf}

\begin{lemma}
The number of four fold-point gluing trees for a polygon
of $n$ vertices is $\Omega(n^2)$ and $O(n^4)$.
\lemlab{4fold.bounds}
\end{lemma}
\begin{pf}
The lower bound is established by a variation on the foldings
of a rectangle to tetrahedra (Fig.~\figref{tetra}).
The idea is to make each of the two conveyor belts 
in a `{\tt I}' structure (Fig.~\figref{I}) realize
$\Omega(n)$ gluings independently.
This can be accomplished by alternating supplementary
angles along the belt at equal intervals.
This is illustrated in Fig.~\figref{zigzag} with angles
$\pi/2$ and $3\pi/2$.  The figure illustrates one possible
folding; the fold points are midpoints of edges.
The tetrahedra produced are the same as that obtained
by folding a rectangle: the ``teeth'' mesh seamlessly.
\begin{figure}[htbp]
\centering
\includegraphics[width=0.6\linewidth]{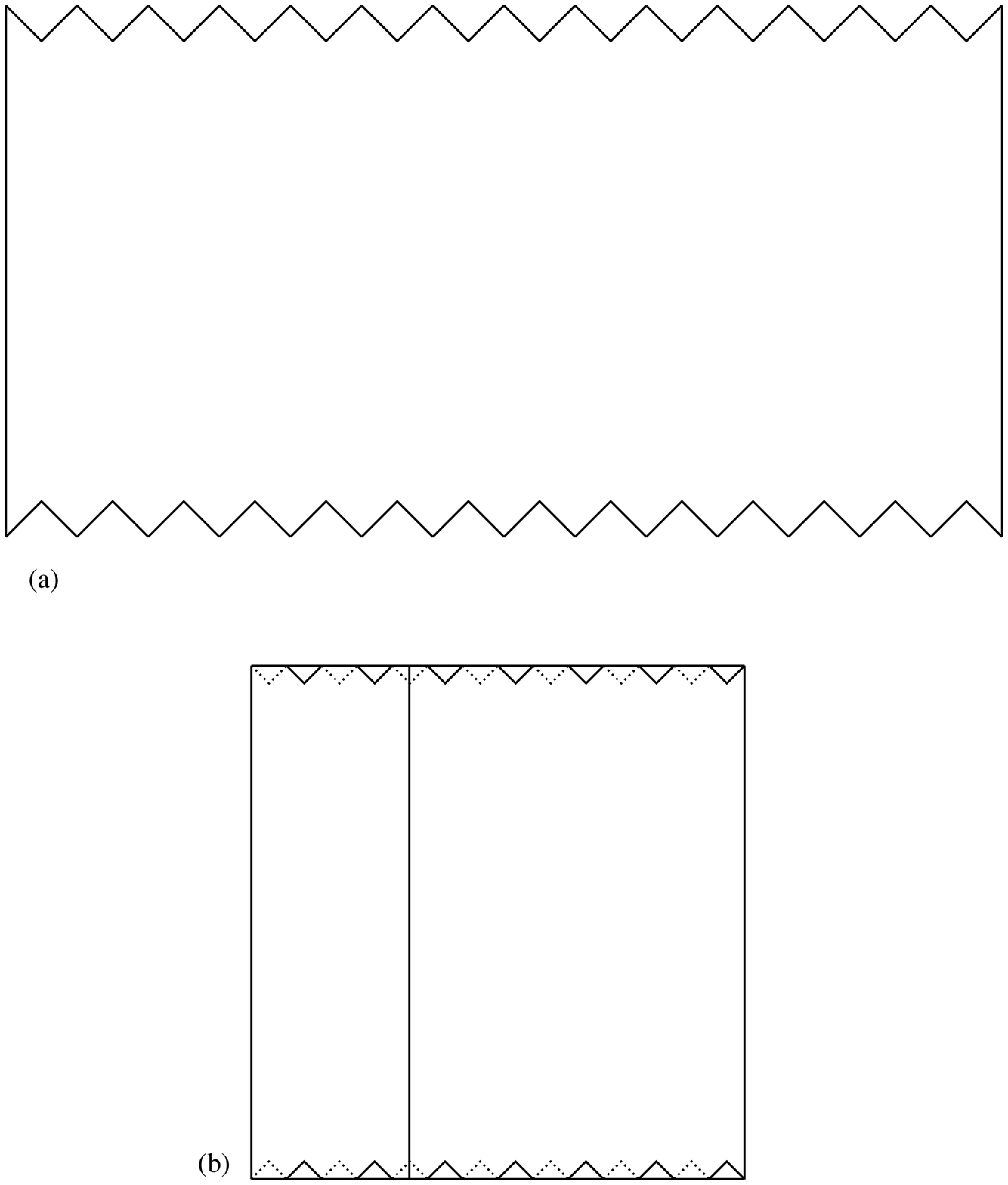}
\caption{(a) Polygon $P$; (b) One four fold-point gluing.
The dashed lines indicate tips of front teeth bent over
and glued behind.}
\figlab{zigzag}
\end{figure}

For the upper bound, Lemma~\lemref{four.4} restricts the 
structures to `{\tt +}' and `{\tt I}'.
\begin{enumerate}
\item `{\tt +}'.
Here we rely on the crude $O(n^4)$ bound determined by the
four vertices, or three vertices and one edge, glued
together to form the interior node of $T_G$.
This fixes the combinatorial type of the gluing, which by
Lemma~\lemref{gluing.tree} determines $T_G$.
\item `{\tt I}'.
Let $a$ and $b$ be the upper and lower nodes of the `{\tt I}'.
There are two cases to consider for the upper node:
\begin{enumerate}
\item $a$ is of type `vv': vertices $v_i$ and $v_j$ glue to
form the belt attachment point.
Then the path from $a$ to $b$ is determined by the requirement that
the curvature must be zero at each point:  the two sides ``zip''
closed from $v_i$/$v_j$ until the first point at which the curvature
is nonzero, which then must be the lower node $b$.

\item $a$ is of type `ve': vertex $v_i$ glues to the interior of edge $e_j$
to form the attachment point.  The ``zipping'' down to $b$ is again
determined, but this takes more argument.
Let $v_{i+1}$ and $v_j$ be the two vertices closest to $a$ on the
path to $b$.  Both of their angles must differ from $\pi$
(because all vertices are essential).
They must glue to one another with an angle sum of $2\pi$
(because the curvature must be zero).  
We want to show that $v_i$ cannot ``slide'' along $e_j$
to another position and still result in an Aleksandrov gluing.
Sliding $v_i$ up
$e_j$ places  $v_{i+1}$ in the interior of $e_j$, and sliding
$v_i$ down places $v_j$ in the interior of $e_i$, in both
cases producing a point of nonzero curvature.  Therefore no sliding
is possible.  Because any respositioning of $v_i$ on $e_j$
can be viewed as such a sliding, no repositioning is possible.
\end{enumerate}
\end{enumerate}
In both cases there are at most $O(n^2)$ choices for the 
constituents glued at $a$.  Together with the $O(n^2)$ bound on
the two belts from Lemma~\lemref{four.n}, this establishes
the claimed $O(n^4)$ bound.
\end{pf}

It seems likely that this lemma could be strengthened: 
\begin{conj}
The number of four fold-point gluing trees for a polygon
of $n$ vertices is $\Theta(n^2)$.
\end{conj}

\subsection{Upper Bound: General Case}
\seclab{Gluing.Upper.General}
We finally are positioned to establish an upper bound on the number
of gluing trees, as a function of the number leaves.

\begin{theorem}
The number of gluing trees with $\l$ leaves for a polygon $P$
with $n$ vertices is $O(n^{2\l-2})$.
\theolab{gluing.upper}
\end{theorem}
\begin{pf}
Let $g(n,\l)$ be the number of gluing trees for $P$ that
have $\l$ leaves.
The proof is by induction on $\l$.
We know from Lemma~\lemref{four.4} that at most four leaves
can be fold-points.
We assume for the general step of the induction that $\l > 4$,
and so there is at least one non-fold-point leaf.
The base cases for $\l \le 4$ will be considered later.

The bound will use one consequence of the angles or curvature of
a gluing (described in this paragraph), and one consequence of the
matching edge lengths of a gluing (described in the next paragraph).
Because a point interior to an edge of $P$ has angle $\pi$, 
a node of degree $d$
of a gluing tree ($d=1, 2, \ldots$) glues together
$d$ vertices of $P$ or $d-1$ vertices and one edge of $P$.
Apart from this, we will use nothing else about the angles of the polygon,
and in fact, our argument will hold more generally for a closed chain
of $n$ vertices, with specified edge lengths.

Given a tree $T_G$ that is not a path, and a leaf $l$, 
define the {\em source\/} of
$l$ as the first node of degree more than 2 along the (unique) path
from $l$ into $T$.  The path in $T_G$ from $l$ to its source is called
the {\em branch\/} of $l$.
For a tree $T_G$ and a non-fold-point leaf corresponding
to polygon vertex $l$, let $s(l)$ be a vertex of $P$ closest to $l$
glued at the source of the leaf.  Note that there must be such a vertex,
since we cannot glue together two points interior to polygon edges at the
source of the leaf.
For example, in Fig.~\figref{tri.tetra}, $s(v_2)$ can be $v_3$ or $v_1$.
Note---this is the single consequence of matching edge lengths referred to
above---that the pair $(l, s(l))$ determines
the portion of $P$'s boundary that is glued together to form the
branch of $l$.
We can simplify $T$ by cutting off $l$'s branch,
resulting in a tree with $\l-1$ leaves.  The corresponding simplification of
$\bP$ is to excise the portion of its chain of length $2d(l, s(l))$
centered at $l$, resulting in a closed chain on at most $n-1$
vertices. Since there are $n$ choices for $l$ and at most
$n$ choices for $s(l)$ we obtain $g(n,\l) \le n^2 g(n-1, \l-1)$.
For the general case there are at most 3 fold leaves, hence:
$g(n,\l) \le n^{2(\l-3)} g(n-(\l-3), 3)$.

Lemmas~\lemref{count.path} and~\lemref{count.Y} established 
the base cases $g(n,2) = O(n^2)$ and $g(n,3) = O(n^4)$.
Substituting, this yields
\begin{eqnarray}
g(n,\l) \le & n^{2(\l-3)} O([n-(\l-3)]^4) \\
         =  & n^{2(\l-3)} O(n^4) \\
         =  & O(n^{2\l-2})
\end{eqnarray}

It remains to handle the case of $\l=4$ leaves.
We will separate into the cases when at least one of these leaves is not
a fold-point leaf, where arguments as above yield $O(n^6)$,
and the case when all 4 vertices are fold leaves.
In this case, Lemma~\lemref{4fold.bounds} establishes
a bound of $O(n^4)$, smaller than that claimed by the lemma.
\end{pf}

Of course because $\l$ could be $\Omega(n)$, there is no contradiction
between this upper bound and the exponential lower bound
in Theorem~\theoref{star}.
We specialize the upper bound to convex polygons in Section~\secref{Gluing.Upper.Convex},
but first we summarize the structural characteristics of
gluing trees we have uncovered.

\subsection{Gluing Tree Characterization}
Our previous results imply that gluing trees are fundamentally
discrete structures, with one or two rolling conveyor belts,
and two such belts only in very special circumstances.
\seclab{Gluing.char}
\begin{theorem}
Gluing trees satisfy these properties:
\begin{enumerate}
\item At any gluing tree point of degree $d \neq 2$,
at most one point of $\bP$ in the interior of an edge may
be glued, i.e., at most one nonvertex may be glued there.
\item At most four leaves of the gluing tree can be fold points,
i.e., points in the interior of an edge of $\bP$.
The case of four fold-point leaves is only possible when
the tree has exactly four leaves, with the combinatorial
structure `{\tt +}' or `{\tt I}'.
\item A gluing tree can have at most two rolling belts.
\item A gluing tree with two rolling conveyor belts must
have the structure `{\tt I}', and result from folding
a polygon that can be viewed as a quadrilateral with two of
its opposite edges replaced by complimentary polygonal paths.
\end{enumerate}
\theolab{gluing.char}
\end{theorem}
\begin{pf}
\begin{enumerate}
\item That $d \neq 2$ points of a gluing tree have at most
one edge-interior points glued is immediate from the
definition of an Aleksandrov gluing, and our insistence
that all vertices are essential.
\item The structure of four fold-point trees was
established in Lemma~\lemref{four.4}.
\item The definition of ``rolling belts'' (p.~\pageref{rolling.belt})
implies four fold points, so the constraints from the previous
item apply.
\item Two rolling belts cannot be accommodated by the `{\tt +}'
structure, which is determined by the four vertices glued
at the central node.  So the tree structure must be `{\tt I}'.
Lemma~\lemref{four.belt} established that the belts are
vertex-free, corresponding to two opposite edges of the
quadrilateral.  The central path of the `{\tt I}'
must be formed by gluing vertices together whose angle sum
is $2\pi$, and they are in this sense complimentary polygonal paths.
\end{enumerate}
\end{pf}

\noindent
Thus a generic gluing tree has one rolling  belt, with
trees hanging off it, 
and one of those trees having a fold-point leaf.
See Fig.~\figref{generic}.
\begin{figure}[htbp]
\centering
\includegraphics[width=0.6\linewidth]{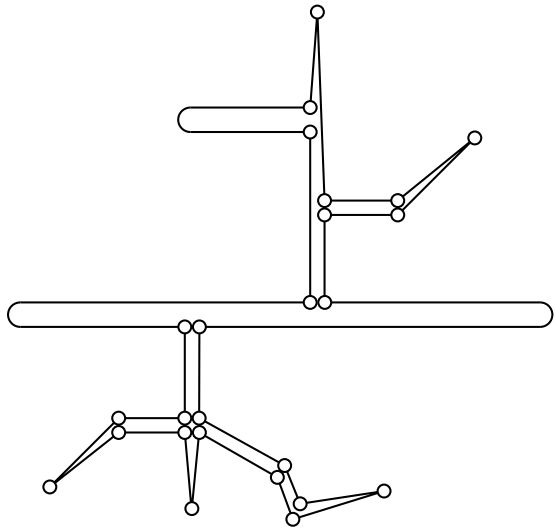}
\caption{A generic gluing tree: three fold-point leaves
(indicated by smooth arcs),
two forming a rolling belt.  Vertices indicated by open circles. }
\figlab{generic}
\end{figure}

\subsection{Upper Bound: Convex Polygons}
\seclab{Gluing.Upper.Convex}
For convex polygons we can prove a polynomial upper bound.
We first handle the special case of $\l=4$.

\begin{lemma}
A convex polygon $P$ may fold to gluing trees of four
leaves only if it is a quadrilateral, a pentagon, or a hexagon;
$P$ may fold to $O(1)$ such gluing trees.
\lemlab{count.+I}
\end{lemma}
\begin{pf}
As in the proof of Theorem~\theoref{cut.comb}, the two conditions
$\g(v) \ge \pi$ and $\sum_v \g(v) = 4\pi$ for the four leaves
$v$ of the tree imply that $\g(v) = \pi$ for each.
This implies that the internal angle at $v$ in $P$ is $\pi$,
which, by our assumption that all vertices are essential,
implies that all four are fold vertices.

Because all available curvature is consumed by the four leaves,
the internal nodes of the gluing tree must be flat.  
If the $T$ has shape `{\tt +}', four vertices
whose angles sum to $2\pi$ join there.
Recalling that the turn angle at each vertex
is $\tau_i = \pi - \a_i$ and that the total turn angle is $2\pi$,
this angle sum implies that $\sum_i \tau_i = 4\pi - 2\pi = 2\pi$,
for the four vertices at the `{\tt +}',
and so the turn angle is completely consumed by these four vertices.
Thus $P$ must be a quadrilateral, and there is just one way
to form the gluing tree.

If $T$ is a `{\tt I}' shape, then 
each of the two internal nodes of the `{\tt I}' are
formed either by gluing together three vertices, or
two vertices and an edge.
For a three-vertex node, the turn angle sum
is $3\pi - 2\pi = \pi$;
for a two-vertex and edge node, the turn angle sum
is $2\pi - \pi = \pi$.
So both nodes together consume of all the $2\pi$ turn angle.
Therefore $P$ has at most six vertices.
The hexagon permits the most groupings of vertices, six;
and so there are at most six gluing trees.
\end{pf}

\noindent
See Fig.~\figref{hex.YY} for an irregular hexagon that folds with
a `{\tt I}' gluing tree.
\begin{figure}[htbp]
\centering
\includegraphics[width=8cm]{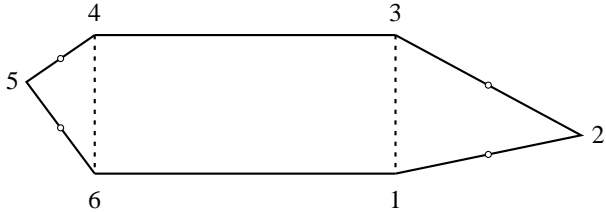}
\caption{A hexagon that folds by gluing together
vertices $\{v_1,v_2,v_3\}$ to form one node,
and $\{v_4,v_5,v_6\}$ to form the other.  Note
the angle sum at the former is $\pi/2+\pi/2$ from the
right angles at $v_1$ and $v_3$,
and $\pi$ from $\triangle(v_1,v_2,v_3)$,
for a total of $2\pi$.
The four fold vertices are marked.
}
\figlab{hex.YY}
\end{figure}

\begin{theorem}
A convex polygon $P$ of $n$ vertices folds to at most
$O(n^3)$ different gluing trees.
\theolab{count.gluing.trees}
\end{theorem}
\begin{pf}
Theorem~\theoref{cut.comb} limits the combinatorial possibilities to
trees with four or fewer leaves.
We have settled each case for $\l \le 4$ earlier:
\begin{description}
\item[$\l=4$] Lemma~\lemref{count.+I}: $O(1)$.
\item[$\l=3$] Lemma~\lemref{count.Y}: $O(n^4)$.
\item[$\l=2$] Lemma~\lemref{count.path}: $O(n^2)$.
\end{description}
We now improve the $\l=3$ case to $O(n^3)$ for convex polygons.
We can tighten the $O(n^4)$ bound with the following two observations:
\begin{enumerate}
\item The two internal angles at the two vertices glued at a type-vve
node must sum to no more than $\pi$.
\item A convex polygon cannot have too many vertices with small angles.
\end{enumerate}
To quantify the second observation, define the {\em turn angle\/}
$\tau_i$ at a vertex $v_i$ with internal angle $\a_i$ to be
$\tau_i = \pi - \a_i$.  For any polyon, we must have
$\sum_i = 2 \pi$.  For a convex polygon, $\tau_i > 0$.
Now suppose $v_i$ and $v_j$ glue at a type-vve node.
Then $\a_i+\a_j \le \pi$, and so
$\tau_i+\tau_j \ge \pi$.
Thus two distinct vv-gluings, involving four different vertices,
already consume the available $2\pi$ turn angle of the polygon.
Call two pairs of vertices {\em disjoint\/} if all four vertices
are different.  The turn angle bound implies
that any polygon can have at
most two disjoint pairs of vertices glued to a type-vve node.
We now show that this implies a $O(n)$ bound on the number of
type-vve nodes.

Construct a bipartite graph $H$ with $n$ nodes for the first vertex
and $n$ nodes for the second vertex of vv-gluings.
Let $(v_i,v_j)$ and $(v_k,v_l)$ be disjoint pairs in vv-gluings,
as depicted in Fig.~\figref{lower.bounds}(b).
Then because there cannot be another pair disjoint from either of
these, every other pair must be incident to one of the
four vertices $v_i,v_j,v_k,v_l$.  This limits $H$ to at most $4n$
edges (and even this bound is loose, for this permits as many as
four disjoint pairs, as is evident in the figure).

Thus there are at most $O(n)$ type-vve nodes.
Repeating the argument that a vv-gluing determines one leg of the `Y'
and Lemma~\lemref{count.path} bounds the remaining path to $O(n^2)$
possibilities, leads to the claimed $O(n^3)$ bound.
\end{pf}

\noindent
We leave open the question of whether this bound is tight.

It is straightforward to list all possible gluing trees for
a given convex polygon with an $O(n^3 \log n)$ time algorithm.
We have implemented such an algorithm, with, however, less than
maximally efficient data structures.

\section{Counting Foldings: Noncongruent Polytopes}
\seclab{Noncongruent.Polytopes}
We have so far been counting the number of different ways to fold
up a given polygon, but have not addressed the question of
whether all these foldings produce distinct polytopes.
There are several notions of what constitutes distinctness.
One natural definition relies on the combinatorial structure
of the polytope, as explored by
Shephard~\cite{s-cpcn-75}.  We will have little to say
on this topic here.
Instead, we will focus on counting noncongruent polytopes.

We have already established in Lemma~\lemref{tetra}
that any rectangle can fold to an uncountably infinite 
number of noncongruent tetrahedra.
We extend this result
in this section to the ``obvious'' fact that 
any convex polygon folds (via perimeter-halving)
to an uncountably infinite number of noncongruent polytopes.
Despite the naturalness of this claim, our inability to determine
the 3D structure of the polytope guaranteed by an Aleksandrov gluing
makes our proof less than satisfactory.
In the absence of any 3D information, we concentrate instead on
the pattern of geodesics between vertices, for of course two
congruent polytopes have the exact same set of geodesics.

\begin{lemma}
A polytope $Q$ resulting from a perimeter-halving fold 
of polygon $P$ has
a countable number of geodesics between any pair of vertices.
\lemlab{geodesics.countable}
\end{lemma}
\begin{pf}
Let $x$ and $y$ be the fold vertices produced by the
perimeter-halving (as in Fig.~\figref{perim.halving}).
We will assign each geodesic a unique integer,
which establishes that there are only a countable number of
them.
The integers are based on a ``layout'' of the
surface of $Q$ in the plane.
Fix $P$ in the plane, and designate it as level-$0$ of the layout.
Around $\bP$ layout $2n$ copies of $P$ (where $P$ has $n$ vertices)
corresponding to the perimeter gluing.
These are level-$1$ $P$ copies of the layout.
This level is illustrated in Fig.~\figref{pent.layout}.
For example, because edge $e_4 = v_4 v_5$ of $\bP$ is glued into
the edge $e_1 = v_1 v_2$ by the perimeter halving, 
a level-$1$ copy of $P$ is placed 
exterior to $e_4$ arranged so that the glued portions of $e_4$ and
$e_1$ match.
There are $2n$ level-$1$ copies of $P$ because the $n$ vertices
around $\bP$ are interspersed by a reversed sequence of the same
$n$ vertices.

Continuing the construction, level-$i$ of the layout is formed by
surrounding each level-$(i{-}1)$ copy of $P$ with $2n$ additional
copies.  Give these copies a ``sequence number'' $j=1,\dots,2n$.
Now every copy of $P$ at level-$i$ in the layout may be assigned a unique
integer by listing the sequence numbers for each level $0,\ldots,i$
and interpreting it as a base-$2n$ number.

It is clear from the layout construction that any geodesic on $Q$
``unrolls'' to a straightline in the layout.
Because we can number the copies of $P$, we can number the geodesics
between any given pair of vertices.
Therefore the number of geodesics is denumerable.
\end{pf}

\noindent
Although this proof is specialized to polytopes formed from perimeter
halving, it would not be difficult to extend it to all polytopes 
formed by gluings including a ``rolling'' fold-point.

\begin{figure}[htbp]
\centering
\includegraphics[width=\linewidth]{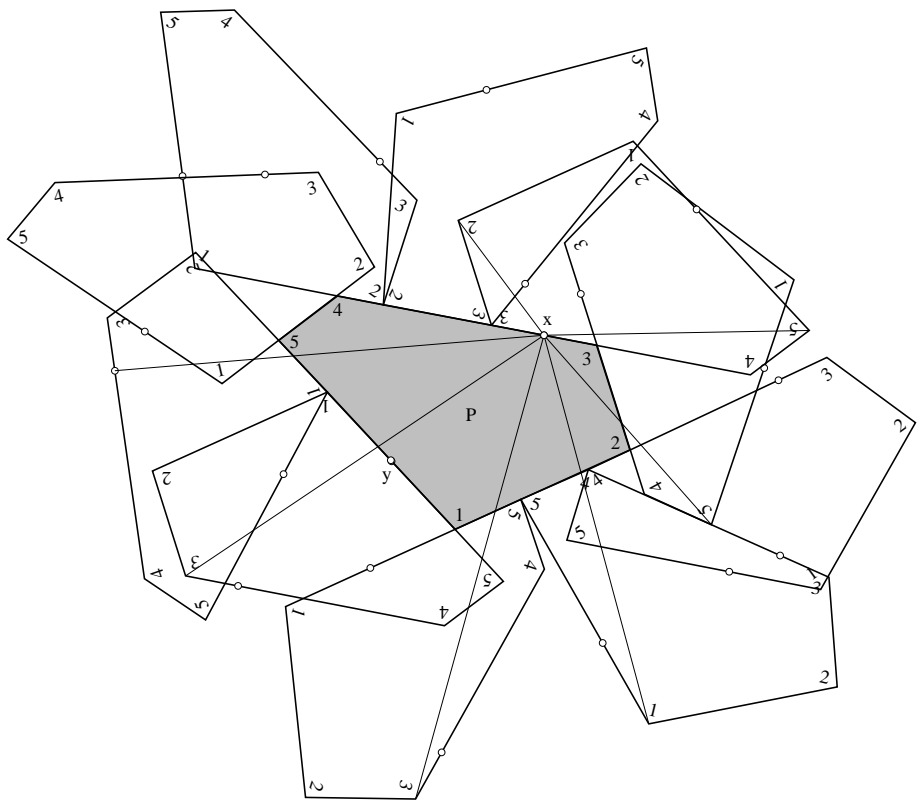}
\caption{Layout of a perimeter-halving folding of a pentagon.
Several geodesics are shown from $x$ to level-$1$ vertices.}
\figlab{pent.layout}
\end{figure}

\begin{theorem}
Any convex polygon $P$ folds, via perimeter halving,
to a uncountably infinite number of
noncongruent polytopes.
\theolab{continuum}
\end{theorem}
\begin{pf}
Select $x$,
a fold point for perimeter halving, 
interior to an edge $e_i = v_i v_{i+1}$ of $P$.
The segment $x v_i \subset \bP$ in level-$0$ of the layout
used in the previous lemma corresponds to a geodesic on $Q_x$.
Now let $x$ vary within some neighborhood $N \subset e_i$;
let $x' \neq x$ be a point in $N$.
The segment $x' v_i$ corresponds to a geodesic on $Q_{x'}$
of a different length.  We use this fact to establish our claim.

Let ${\cal Q} = \{Q_{x'} \;:\; x' \in N \}$ be the
set of all the polytopes produced as $x$ varies over the neighborhood.
Assume, for the purposes of contradiction, that the number
of distinct, noncongruent polytopes in $\cal Q$ is denumerable:
$Q_1, Q_2, \ldots$.
By Lemma~\lemref{geodesics.countable}, each has a countable
number of geodesics: a pair of numbers suffice to uniquely identify them.
Thus the total number of distinct lengths of geodesics represented by
all these polytopes is denumerable.
But this contradicts the nondenumerable number of lengths of
segments $|x' v_i|$ for $x' \in N$.
Therefore the number of noncongruent polytopes in $\cal Q$ is
nondenumerable.
\end{pf}

Although this theorem establishes the result even for regular
polygons, there is much more to say about the structure of
the polytopes that can be folded from regular polygons.
We explore this in Section~\secref{Regular}.

\section{Counting Unfoldings: Cut Trees}
\seclab{Unfolding.Cut.Trees}
In this section we explore unfolding from the point of view of
cut trees.  The general situation is that we are given one
polytope $Q$ of $n$ vertices, and we would like to know
how many different ways it can be cut and unfolded to a polygon.
We start with some straightforward observations before
proving enumeration bounds.

First, every polytope admits at least the $n$ cut trees
provided by the star unfolding~\cite{ao-nsu-92}, one with each
vertex as source.
So in particular, every polytope unfolds to at least one polygon.
(As we mentioned in the Introduction, the corresponding
question for edge-unfoldings remains open.)

Second, because we permit arbitary polygonal paths between the
nodes of a cut tree (Section~\secref{cut.glue.trees}),
there is no upper bound on the number of polygon vertices in
potential unfoldings of a given polytope.  This might lead one
to wonder if \emph{any} polygon (of the appropriate area) 
can be unfolded from
a given polytope.  The answer is {\sc no}, as is easily
established by this lemma.

\begin{lemma}
Every polygon $P$ cut from $Q$ must have at least two vertices whose
interior angles are of the form $2\pi - \g_i$ for some $i=1,\ldots,n$, 
where $\g_i$ are the
curvatures of the vertices of $Q$.
\lemlab{not.all.polygons}
\end{lemma}
\begin{pf}
Let the $n$ vertices of $Q$ have curvatures $\g_i$, $i=1,\ldots,n$.
The cut tree $T_C$ must
have at least two leaves,
and by Lemma~\lemref{cut.tree} these leaves must be vertices of $Q$.
Say they coincide with the vertices of
curvatures $\g_1$ and $\g_2$.  
Then any polygon $P$ that unfolds from $T_C$ must have
two vertices with interior angles $2\pi - \g_1$ and $2\pi - \g_2$.
\end{pf}

\noindent
So let $P$ be a polygon with no interior angle equal to $2\pi - \g_i$
for $i=1,\ldots,n$.  Then $P$ cannot be cut from $Q$.

\subsection{Lower Bound: Exponential Number of Unfoldings}
In this section we provide an exponential lower bound.

\begin{theorem}
There is a polytope $Q$ of $n$ vertices
that may be cut open with exponentially many {\rm ($2^{\Omega(n)}$)}
combinatorially distinct
cut trees,
which unfold to exponentially many geometrically distinct simple polygons.
\theolab{volcano}
\end{theorem}
\begin{pf}
$Q$ is a truncated cone, as illustrated in 
Fig.~\figref{volcano.0}:
\begin{figure}[htbp]
\centering
\includegraphics[width=0.5\linewidth]{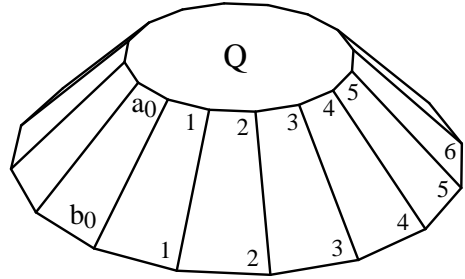}
\caption{
Polytope $Q$.
}
\figlab{volcano.0}
\end{figure}
the hull of
two regular $n$-gons of different radii, lying in parallel planes
and similarly oriented.
We call this the {\em volcano\/} example.
We require that $n$ be even; in the figure, $n=16$.
Label the vertices on the top face $a_0,\ldots,a_{n-1}$
and $b_0,\ldots,b_{n-1}$ correspondingly on the bottom face.
The ``base'' cut tree, which we notate as $T_{0000000}$,
unfolds $Q$ as shown in Fig.~\figref{volcano.1}.
$T_{0000000}$ consists of a path on the top face
$(a_0,a_1,\ldots,a_{n-1})$ supplemented by arcs
$(a_i,b_i)$ for all $i=0,\ldots,n-1$.
The polygon $P$ produced consists of the base face,
$n$ attached trapezoids $(b_i,b_{i+1},a_{i+1},a_i)$,
with the top face attached to $a_{n-1} a_0$.

\begin{figure}[htbp]
\centering
\includegraphics[width=\linewidth]{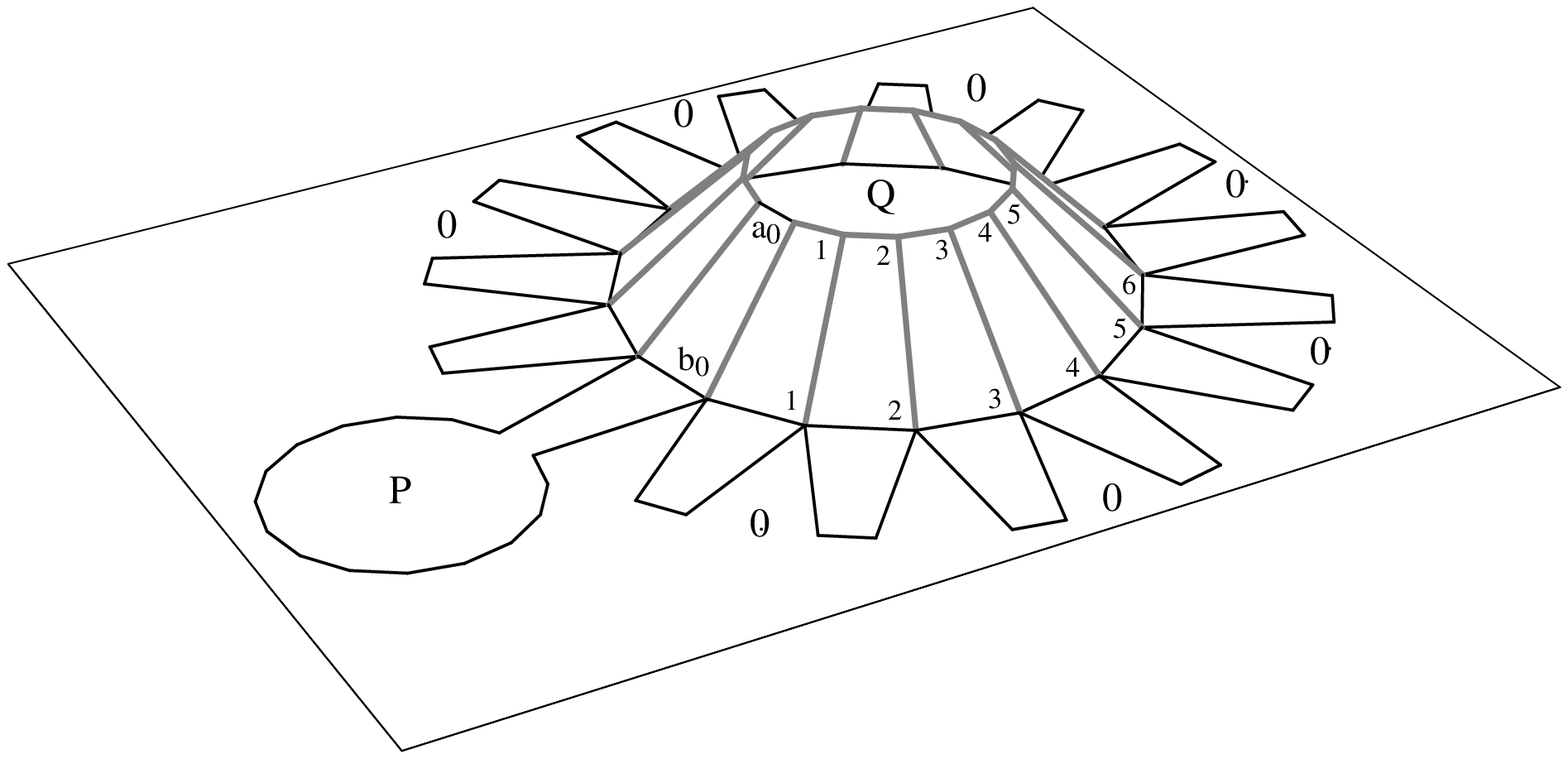}
\caption{
Unfolding via shaded cut tree $T_{0000000}$.
}
\figlab{volcano.1}
\vspace{5mm}
\centering
\includegraphics[width=\linewidth]{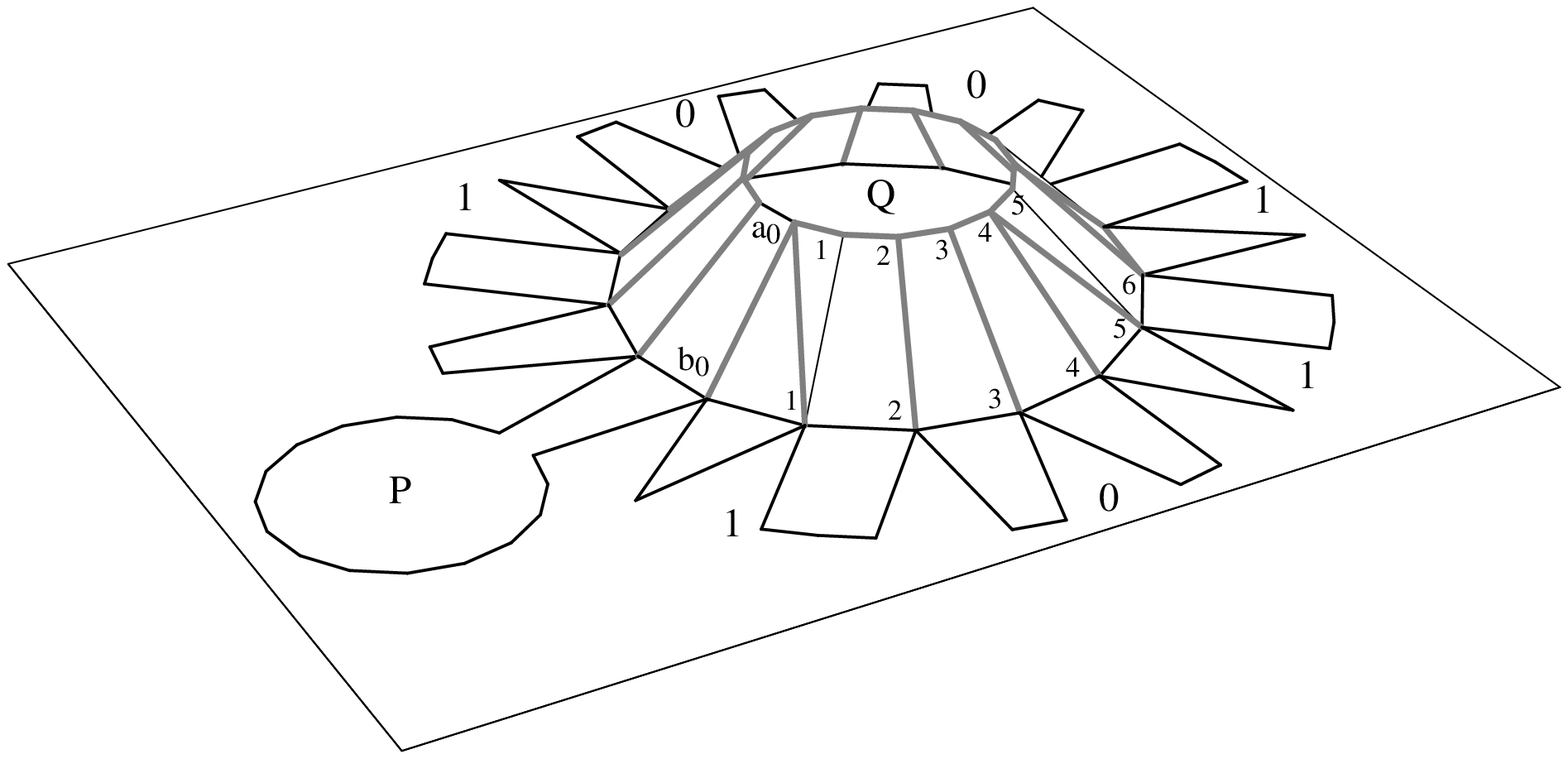}
\caption{
Unfolding via shaded cut tree $T_{1001101}$.
}
\figlab{volcano.2}
\end{figure}
Define a cut tree $T_{m_{(n-1)/2} \cdots m_2 m_1 m_0}$,
where $m_i$ are the digits of a binary number
of $n/2 - 1$ bits,
as an alteration of the base tree $T_{0\cdots0}$ as follows.
If $m_i = 1$, then the arc
$(a_{2i+1},b_{2i+1})$ is deleted and replaced by
$(a_{2i},b_{2i+1})$.
If $m_i = 0$, then the arc $(a_{2i+1},b_{2i+1})$ is used as in  $T_{0\cdots0}$.
Thus the cut tree $T_{1001101}$ shown in Fig.~\figref{volcano.2}
replaces 
$(a_1,b_1)$ with $(a_0,b_1)$ because $m_0 = 1$,
$(a_5,b_5)$ with $(a_4,b_5)$ because $m_2 = 1$,
and so on.

There are $2^{n/2 - 1} = 2^{\Omega(n)}$ cut trees.

It is clear by construction that all these cut trees lead
to simple polygon unfoldings.
It only remains to argue that each leads to a distinct polygon,
not congruent to any other.  This is not strictly true for $Q$ as
defined, for any bit pattern leads to a $P$ that is congruent
(by reflection) to the polygon obtained from the reverse of the bit
pattern.  However, it is a simple matter
to introduce some asymmetry, by, for example, lengthening edge
$a_{n-1} a_0$ slightly.  Then all cut trees lead to distinct
polygons.
\end{pf}

\noindent
A simpler example is a drum, the convex hull of two regular polygons
in parallel planes.  Because some of the unfoldings used in the
above proof overlap, there is a bit more argument needed to establish
the exponential lower bound.

\begin{figure}[htbp]
\centering
\includegraphics[width=\linewidth]{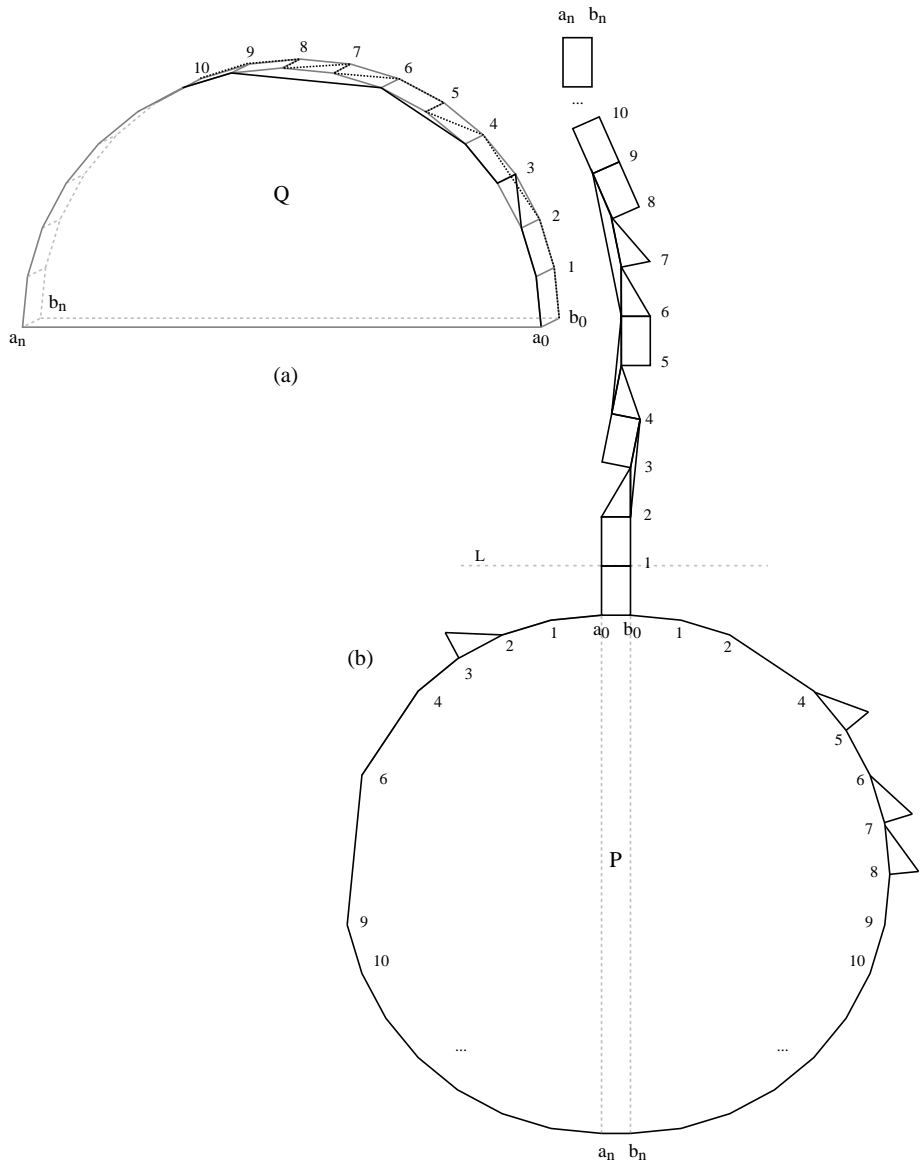}
\caption{
(a) Polytope $Q$,
cut tree $T_{\cdots 0 0 2 2 0 2 0 1 0 0}$.
The $a$-path is shown solid, the $b$-path dashed.
(b) Unfolding to polygon $P$. $L$ strictly separates the head
from the tail.
}
\figlab{sperm}
\end{figure}

Even restricting the cut tree to a path permits
an exponential number of unfoldings:

\begin{theorem}
There is a polytope $Q$ of $n$ vertices
that may be cut open with exponentially many
{\rm ($2^{\Omega(n)}$)}
combinatorially distinct
cut trees, all of which are paths,
which unfold to exponentially many geometrically
distinct simple polygons.
\theolab{sperm}
\end{theorem}
\begin{pf}
$Q$ is formed by pasting two halves of a regular $2n$-gon
together to form a semicircle approximation with some 
small thickness $w > 0$.
Label the vertices on the front face $a_0,\ldots,a_n$
and $b_0,\ldots,b_n$ correspondingly on the back face,
as illustrated in
Fig.~\figref{sperm}(a).
Let $\a = 2\pi/n$ be the turn angle at each vertex $a_i$ (and at $b_i$),
i.e., the angle $\a = \pi - \angle (a_{i-1},a_i,a_{i+1})$.
(In the figure, $\a = \pi/32 \approx 11^\circ$.)
We specify a series of cut trees $T_m$, where $m$ is an
$n$-digit base-$3$ integer $m_n \cdots m_2 m_1$,
with the following interpretation.
$T_{0 \cdots 00}$ is the ``base'' cut tree on which
all others are variations:
\begin{equation}
T_{0 \cdots 00} = (a_0,a_1,\ldots,a_n,b_n,b_{n-1},\ldots,b_1,b_0)
\end{equation}
Note that $T_{0 \cdots 00}$ is a path, as are all the $T_m$.
We call the half of the path on the front face the $a$-path,
and that on the back the $b$-path.
The unfolding $P$ determined by $T_{0 \cdots 00}$ is a regular
$2n$-gon, fattened by a strip $(a_0, b_0, b_n, a_n)$ of width $w$ down
its middle, with a ``tail'' of $n$ rectangles attached to edge $a_0 b_0$.
If $|a_i a_{i+1}| = h$, then each rectangle is $w \times h$.

In cut tree $T_{m_n \cdots m_2 m_1}$
the index $m_i$ is $1$ if the $a$-path deviates to touch $b_i$
on the back face
via the path $(\ldots,a_{i-1},b_i,a_i,a_{i+1},\ldots)$,
and the index $m_i$ is $2$ if the $b$-path similarly deviates
to include $a_i$ on the front face
via the path $(\ldots,b_{i-1},a_i,b_i,b_{i+1},\ldots)$.
In both cases, the opposite path skips the vertex deviated to:
if $m_i=1$, the $b$-path skips $b_i$ by shortcutting on the back face, and
if $m_i=2$, the $a$-path skips $a_i$ by shortcutting on the front face.
Fig.~\figref{sperm}(a)
illustrates $T_{m_n \cdots 0 0 2 2 0 2 0 1 0 0}$,
with $a$-path
\begin{equation}
(a_0,a_1,a_2,b_3,a_3,a_4,a_6,a_9,a_{10},\ldots,a_n)
\end{equation}
and $b$-path
\begin{equation}
(b_0,b_1,b_2,b_4,a_5,b_5,b_6,a_7,b_7,a_8,b_8,b_9,b_{10},\ldots,b_n)
\end{equation}
Note that when $m_i \neq 0$, the rectangle bounded between $a_{i-1} b_{i-1}$
and $a_i b_i$ is crossed by an $ab$-diagonal.
We insist that $m_1 = 0$, so that the cut tree starts
with an uncrossed rectangle $(a_0, b_0, b_1, a_1)$.
Finally, the edge $a_n b_n$ is included in $T_m$, so that
it is a path from
$a_0$ to $a_n$ to $b_n$ and returning to $b_0$.
The digits $m_n \cdots m_2$ are each free to
be any one of $\{0,1,2\}$.
Thus there are an exponential number of combinatorially
distinct $T_m$:  $3^{n-1}$.
We return below to the issue of how many of these lead to geometrically
distinct unfoldings.

It should be clear by construction that $T_m$ spans the vertices.
To show that it is a tree, we need to argue that it is non-self-intersecting.
This is again clear by construction, for each nonzero $m_i$ uses
a diagonal in the rectangle prior to $a_i b_i$, and because
$m_i$ has only one value, no such rectangle has both diagonals used.
Together with the shortcutting that prevents the $a$- and $b$-paths
from touching the same vertex, it follows that $T_m$ is indeed a tree;
so
it is a legitimate cut tree.
Thus it unfolds to a single piece.
It only remains to show that this unfolding is a simple
polygon, i.e., it avoids overlap.

This is obvious
for  $T_{0 \cdots 00}$, as mentioned previously.
For general $T_m$, consider the layout of the unfolding $P$ that
places $a_0 b_0$ horizontally, as in Fig.~\figref{sperm}(b).
Let $L$ be the horizontal line through $a_1 b_1$;
this segment is necessarily horizontal because we stipulated
that  $m_1 = 0$.  We will argue that $L$ strictly separates the 
{\em tail\/} of $P$ (the portion attached above $a_1 b_1$)
from its {\em body\/} (the portion attached below $a_1 b_1$).

First, it is clear that the body unfolds without overlap.
For it is simply truncations (due to path shortcuttings) of
halves of a regular polygon glued to either side of the rectangle
$(a_0, b_0, b_n, a_n)$, with attached triangle ``spikes''
for each nonzero $m_i$.  None of these spikes can overlap,
even when adjacent, for their length-$w$ edge juts out
orthogonal to their length-$h$ edge glued to the body
(see the body image of $b_7$ in Fig.~\figref{sperm}(b)).

The tail consists of $h \times w$ rectangles, or half-rectangles,
glued end-to-end, with turns to the right by $\a$ for every
$m_i=1$ digit, and turns to the left by $\a$ for every $m_i=2$
digit.  Thus, $T_{\cdots 0 0 2 2 0 2 0 1 0 0}$ in (b)
of the figure turns right once and left three times.
Because there are at most $n-1$ nonzero digits, the tail
can turn at most $n-1$ times.  Because $\a$ is the turn angle
of a regular $2n$-gon, it takes $n$ turns of $\a$ to turn
a full $\pi$.  Thus the tail turns strictly less than $\pi$,
and so cannot return to line $L$.  Thus the tail remains
strictly above $L$.
Choosing $w < h$ guarantees that no body spike protrudes vertically
as much as $h$ above $a_0 b_0$; so the body remains strictly below $L$.

It remains to argue that the tail does not self-intersect.
But this follows from the same turn argument above.
By construction, there are no local overlaps between two
adjacent tail rectangles or half-rectangles.
Thus the only overlap conceivable would result from the tail
curling back to overlap itself.
Choosing $w \ll h$ makes the tail essentially a series of segments
of length $h$, with attached pieces of the regular polygon
clipped by shortcutting.
For the tail segments to curl back and overlap
would require a
total turn by at least $\pi$,
contradicting the bound
on the sum of $\a$'s.

Finally, we turn to the question of how many of the 
$3^{n-1}$ combinatorially distinct $T_m$ lead to geometrically
distinct (noncongruent) $P$.
Let $x$ and $y$ be two base-$3$ numbers,
and let $S(x)$ be the base-$3$ number obtained by changing each
$1$-digit in $x$ to a $2$, and each $2$-digit in $x$ to a $1$.
(For example, $S(1021) = 2012$.)
Then if $S(x) = y$, $T_x$ and $T_y$ lead to congruent $P$,
in that $P_y$ is the reflection of $P_x$ about a vertical line
(in the layout used above).

Although we could easily ensure noncongruency for
all $T_m$ by altering $Q$ to be
less symmetric, we opt here for a counting argument.
Let $x$ be a base-$3$ number.  Define $B(x)$ to be the binary
number obtained by changing each $2$-digit in $x$ to a $1$.
(For example, $B(2012) = 1011$.)
Now it should be clear that for any two
base-$3$ numbers $x$ and $y$, if $B(x) \neq B(y)$, then
$P_x$ is noncongruent to $P_y$.  For then the pattern of spikes
on the body are different in $P_x$ and $P_y$.
Thus, among the $3^{n-1}$ combinatorially distinct $P$,
there are at least $2^{n-1}$ geometrically distinct $P$.
\end{pf}

\subsection{Lower Bound: Convex Unfoldings}
It seems possible that the exponential lower bound holds
even in the case of convex unfoldings,
via an example similar to that used in Fig.~\figref{sperm}.
\begin{conj}
There is a polytope with an exponential number of
convex unfoldings.
\end{conj}

\noindent
This represents the only `?' in Table~\tabref{results}.

\subsection{Upper Bound}

\begin{theorem}
The maximum number of edge-unfolding cut trees of a polytope of $n$
vertices is $2^{O(n)}$,
and the maximum number of arbitary cut trees
$2^{O(n^2)}$.
\end{theorem}
\begin{pf}
For edge unfoldings, the bound depends on the number
of spanning trees of a polytope graph.
We may obtain a bound here as follows.%
\footnote{
	We thank B. McKay [personal communication, Jan.~2000] for guidance here.
}
First, 
triangulating a planar graph only increases the number of spanning trees,
so we may restrict attention to
triangulated planar graphs.
Second, it is well known that 
the number of spanning trees of a connected planar graph is the same as
the number of spanning trees of its dual.
So we focus just on $3$-regular (cubic) planar graphs.
Finally, a result of McKay~\cite{m-strg-83} proves an 
upper bound of $O((16/3)^n / n)$ on the number of spanning trees
for cubic graphs.
This bound is $2^{O(n)}$.

For arbitrary cut trees, the underlying graph 
might conceivably have a quadratic number of edges,
which leads to the bound $2^{O(n^2)}$.
(Note that our definition of cut tree in
Section~\secref{Cut.Trees} would not count different polygonal
paths between two vertices as distinct arcs of $T_C$.)
\end{pf}

\section{Counting Unfoldings: Noncongruent Polygons}
\seclab{Noncongruent.Polygons}
We have already seen in Theorem~\theoref{volcano}
that one polytope can have an exponential number of noncongruent
polygon unfoldings.
In fact the possibilities range from $0$ to $\infty$,
even for convex unfoldings,
as this simple
counterpart of Theorem~\theoref{continuum} shows:
\begin{figure}[htbp]
\centering
\includegraphics[width=8cm]{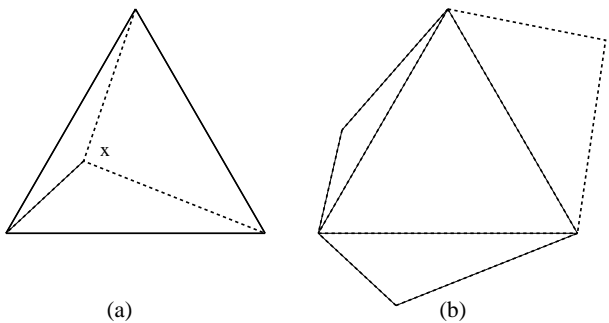}
\caption{Unfolding a flat triangle (a) to a convex polygon (b).}
\figlab{tri.x}
\end{figure}

\begin{theorem}
Although some polytopes
unfold to a nondenumerable number of noncongruent
convex polygons, others have only a finite number of
convex unfoldings.
\end{theorem}
\begin{pf}
For the former claim, consider a doubly-covered equilateral
triangle.  Choose any point $x$ interior to the top face,
as shown in Fig.~\figref{tri.x}(a).  This leads to a `{\tt Y}' cut tree
that unfolds to a convex polygon~(b) for every choice of $x$.
All these polygons have different angles, and so are noncongruent.

The second claim of the theorem is trivially satisfied by polytopes
with zero convex unfoldings.
To establish it for a polytope that has at least one convex
unfolding is more difficult, and 
we only sketch a construction.
Consider the doubly-covered trapezoid shown in  Fig.~\figref{trap}.
It has just two sharp vertices, $v_1$ and $v_4$, and so,
by Theorem~\theoref{cut.comb}, the cut tree must be a path
connecting those vertices.  The path $(v_1,v_2,v_3,v_4)$
unfolds to a convex polygon.
Now consider a geodesic that starts with the segment $v_1 v_3$
as illustrated.  As in the proof of Lemma~\lemref{not.two.sharp},
this geodesic will either hit $v_4$ directly, in which case it is not a
valid cut tree because $v_2$ is not spanned, or
it spirals around the trapezoid and self-crosses.
We will not prove this claim.
\end{pf}

\begin{figure}[htbp]
\centering
\includegraphics[width=8cm]{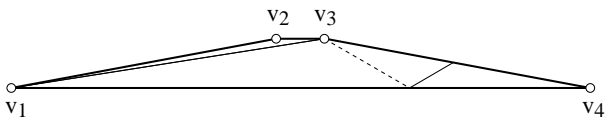}
\caption{A geodesic on a trapezoid from $v_1$ through $v_3$.}
\figlab{trap}
\end{figure}

\section{Folding Regular Polygons}
\seclab{Regular}
In this section we study folding \emph{regular} polygons
of $n$ vertices.
Because all polygon vertices have the same interior angle $\t = (n-2)\pi/n$,
only a limited variety of different polytope vertex curvatures
may be created.  
We find, not surprisingly, that this leads to a limited
set of possibilities:
in general,
only one ``class'' of nonflat polytopes can be produced.
This is established in Lemma~\lemref{ngt6}.

Let $\a_k$, $k \ge 1$, be the curvature at a polytope vertex
formed by gluing $k$ $P$-angles of $\t$ together,
and $\b_k$, $k \ge 0$, be the curvature at a vertex formed
by gluing $k$ angles to a point interior to an edge of $\bP$.
The next lemma details the possible $\a_k$ and $\b_k$ values
achievable.

Throughout this section we will find that the situation is more
uniform for $n > 6$ than it is for small $n$.

\begin{lemma}
For $n > 6$, only four vertex curvatures can be obtained
by folding a regular $n$-gon $P$; 
for $n \le 6$, additional curvature values are
possible.
More precisely,
for all $n$, these four curvature values are always achievable:
\begin{itemize}
\item $\a_1 = \pi(1+2/n)$.
\item $\a_2 = \pi(4/n)$.
\item $\b_0 = \pi$.
\item $\b_1 = \pi(2/n)$.
\end{itemize}
The additional values possible for $n \le 6$
are detailed in Tables~\tabref{alpha} and ~\tabref{beta}.
\lemlab{curvature.values}
\end{lemma}
\begin{pf}
\begin{enumerate}
\item $\a_1 = 2\pi - \t = 2\pi - (n-2)\pi/n = \pi(1+2/n)$.
This vertex is a leaf of the gluing/cut tree.
We call this a {\em zipped\/} vertex, for $\bP$ is
``zipped shut'' at the vertex.
\item $\a_2 = 2\pi - 2\t = 2\pi - 2(n-2)\pi/n =\pi(4/n)$.
This vertex is a degree-$1$ node of the gluing tree.
\item $\b_0 = \pi$.
This is a fold vertex, when nothing is glued to an edge of $\bP$,
and therefore a leaf of the gluing tree.
\item $\b_1 = 2\pi - [\pi + \t] = 2\pi - [\pi + (n-2)\pi/n] = \pi(2/n)$.
This is a degree-$1$ node of the gluing tree.
\end{enumerate}
The additional possibilities for $n \le 6$ are as follows.
$\a_3$ is possible for all $n \le 6$;
$\a_4$ is possible only for $n=4$;
and no other $\a_k$ is possible.
See Table~\tabref{alpha}.

For $n=3$, $\b_2, \b_3$, and for $n=4$, $\b_2$, are all possible.
See Table~\tabref{beta}.

Explicit computation
shows that all higher values of $k$ lead to nonconvex vertices, 
whose total face
angle exceeds $2\pi$ and so which have negative curvature.
\end{pf}

\begin{table}[htbp]
\begin{center}
\begin{tabular}{| c | c | c | c | c |}
        \hline
$\a_k$ & \multicolumn{4}{c|}{$k$}
        \\ \hline
\mbox{} & 1 & 2 & 3 & 4
        \\ \hline \hline
\rule[-6pt]{0pt}{17pt}3
        & $\frac{5}{3}$
        & $\frac{4}{3}$
        & $1$
        & \mbox{}
        \\ 
\rule[-6pt]{0pt}{17pt}4
        & $\frac{3}{2}$
        & $1$
        & $\frac{1}{2}$
        & $0$
        \\ 
\rule[-6pt]{0pt}{17pt}5
        & $\frac{7}{5}$
        & $\frac{4}{5}$
        & $\frac{1}{5}$
        & \mbox{}
        \\ 
\rule[-6pt]{0pt}{17pt}6
        & $\frac{4}{3}$
        & $\frac{2}{3}$
        & $0$
        & \mbox{}
        \\ \cline{1-3}
\rule[-6pt]{0pt}{17pt}$n$
        & $1+\frac{2}{n}$
        & $\frac{4}{n}$
        & \multicolumn{2}{c|}{\mbox{}}
        \\ \hline
\end{tabular}
\end{center}
\caption{Possible $\a_k$ curvature values, in units of $\pi$.}
\tablab{alpha}
\end{table}

\begin{table}[htbp]
\begin{center}
\begin{tabular}{| c | c | c | c | c |}
        \hline
$\b_k$ & \multicolumn{4}{c|}{$k$}
        \\ \hline
\mbox{} & 0 & 1 & 2 & 3
        \\ \hline \hline
\rule[-6pt]{0pt}{17pt}3
        & $1$
        & $\frac{2}{3}$
        & $\frac{1}{3}$
        & $0$
        \\ 
\rule[-6pt]{0pt}{17pt}4
        & $1$
        & $\frac{1}{2}$
        & $0$
        & \multicolumn{1}{c|}{\mbox{}}
        \\ \cline{1-3}
\rule[-6pt]{0pt}{17pt}$n$
        & $1$
        & $\frac{2}{n}$
        & \multicolumn{2}{c|}{\mbox{}}
        \\ \hline
\end{tabular}
\end{center}
\caption{Possible $\b_k$ curvature values, in units of $\pi$.}
\tablab{beta}
\end{table}

Let $a_i$ and $b_i$ be the number of polytope vertices of curvature
$\a_i$ and $\b_i$ respectively, formed by folding
a regular $n$-gon $P$.
Of course $a_i$ and $b_i$ are nonnegative integers,
but there are additional significant restrictions imposed by
the requirement that the total curvature be $4 \pi$:
\begin{equation}
\sum_{i=1} a_i \a_i + \sum_{i=0} b_i \b_i = 4 \pi \;. \eqlab{4pi}
\end{equation}
We now explore the implications of this constraint,
separately for $n > 6$ and for $n \le 6$.
Note that our notation implies that
\begin{equation}
\sum_{i=1} a_i i + \sum_{i=0} b_i i =  n \;, \eqlab{indices}
\end{equation}
because the subscripts on $\a$ and $\b$ indicate the number of
vertices involved in the gluing.

\begin{figure}[htbp]
\centering
\includegraphics[width=0.5\linewidth]{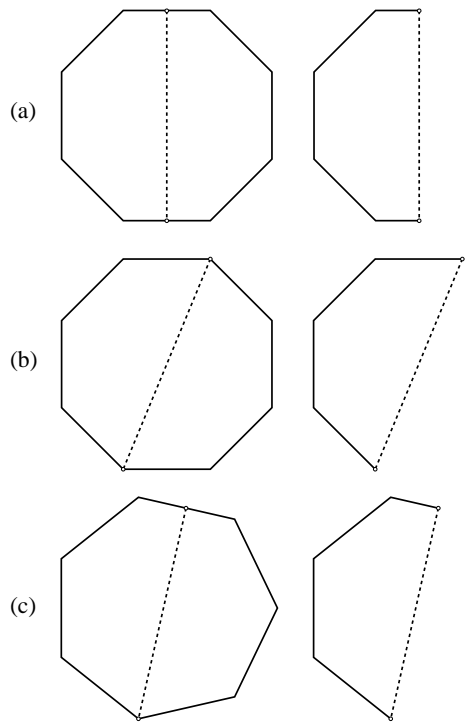}
\caption{(a,b) Flat foldings of an $n$-gon, $n$ even
($n=8$).
(c) Flat folding of an $n$-gon, $n$ odd ($n=7$).}
\figlab{flat.cases}
\end{figure}

Now we prove that
perimeter-halving is the only possible kind of folding for $n>6$.

\begin{lemma}
For all $n \ge 3$,
regular $n$-gons fold via
perimeter-halving, using path gluing trees,
to two classes of polytopes:
\begin{enumerate}
\item A continuum of ``pita'' polytopes of $n+2$ vertices.
\item One or two flat, ``half-$n$-gons'':
\begin{enumerate}
\item $n$ even: Two flat polytopes, of $\frac{n}{2}+2$ and $\frac{n}{2}+1$ vertices.
\item $n$ odd: One flat polytope, of $\frac{n+1}{2}+1$ vertices.
\end{enumerate}
\end{enumerate}
For $n > 6$, these are the only foldings possible of
a regular $n$-gon.
\lemlab{ngt6}
\end{lemma}
\begin{pf}
A perimeter-halving fold produces a path gluing tree.
This has two leaves and all other nodes internal.
From
Lemma~\lemref{curvature.values},
the only two curvatures can be leaves:
$\{ \a_1, \b_0 \}$;
and only two can be degree-$1$ nodes:
$\{ \a_2, \b_1 \}$.
Moreover, these are the
only curvatures possible
for $n > 6$.  Thus Eq.~\eqref{4pi}
reduces to
\begin{equation}
a_1 \a_1 + a_2 \a_2 + b_0 \b_0 + b_1 \b_1 = 4 \pi \;.
\end{equation}
Substituting the curvature values from 
Lemma~\lemref{curvature.values} and solving for $n$ yields
\begin{equation}
n = \frac{2(a_1 + 2 a_2 + b_1)}
        {4-(a_1 + b_0)} \; \eqlab{4pi.n}
\end{equation}
Because
only $\a_1$ and $\b_0$ are leaf vertex curvatures,
we must have $a_1 + b_0 \ge 2$.
The requirement that the denominator of Eq.~\eqref{4pi.n}
be positive yields  $a_1 + b_0 < 4$.
Therefore we know that  $a_1 + b_0 \in \{2, 3\}$.
We now show that the case $a_1 + b_0 = 3$ is not possible when $n > 6$.

As both $a_1$ and $b_0$ count leaves, a tree formed
with  $a_1 + b_0 = 3$ must have at least three leaves.
By Theorem~\theoref{cut.comb}, because $n \neq 4$,
it cannot have more than three leaves.
So it has exactly
three leaves, and has the combinatorial structure of a `{\tt Y}'.
The interior node must be formed by gluing three distinct points of
$\bP$ together (by Lemma~\lemref{cut.tree}(4)).
This corresponds to curvatures
$\a_k$, $k \ge 3$, or $\b_k$, $k \ge 2$.
But Lemma~\lemref{curvature.values} shows that none of these
are possible for $n > 6$.
(Note, for later reference, that for $n \le 6$, these possibilities
will need consideration.)

Therefore we must have  $a_1 + b_0 = 2$.
Therefore the gluing tree must be a path for $n > 6$,
and the folding must be a perimeter-halving folding.
We now explore the three possible solutions to $a_1 + b_0 = 2$.
\begin{description}
\item[Case $a_1=0$, $b_0=2$.]
The two leaves are both folds at interior points of edges
of $\bP$, a perimeter-halving folding similar to that
previously illustrated in 
Fig.~\figref{perim.halving}.
If neither fold point $x$ and $y$
is the midpoint of its edge, then
no pair of vertices glue together, so $a_2 = 0$
and therefore $b_1 = n$.
This produces a continuum of polytopes $Q_x$ of $n+2$ vertices.
We call these {\em pita polytopes}
(Fig.~\figref{pita}), and will study them in
Section~\secref{pita} below.

Suppose one fold point $x$ is at an edge midpoint.
If $n$ is even, then $y$ is also at a midpoint,
and $P$'s vertices are glued in pairs.
Therefore $a_2 = n/2$ and $b_1 = 0$.
The polytope is a flat half-$n$-gon of $n/2+2$ vertices.
See Fig.~\figref{flat.cases}(a).
If $n$ is odd, then $y$ must be at a vertex.  This means that
$a_1 \neq 0$, and this case does not apply.

\item[Case $a_1=2$, $b_0=0$.]
Both leaves are at vertices, and so $n$ must be even.
All other vertices are glued in pairs, so
$a_2 = (n-2)/2$ and $b_1 = 0$.
The folding produces a flat half-$n$-gon
of $n/2+1$ vertices.
See Fig.~\figref{flat.cases}(b).

\item[Case $a_1=1$, $b_0=1$.]
One vertex is zipped to a leaf; half the perimeter around
is a fold vertex.  This implies that $n$ is odd.
All other vertices are glued in pairs, so
$a_2 = (n-1)/2$ and $b_1 = 0$.
The folding produces a flat half-$n$-gon
of $(n-1)/2+2$ vertices.
See Fig.~\figref{flat.cases}(c).
\end{description}
The details derived above are gathered into
Table~\tabref{flat.cases},
and the flat foldings illustrated in Fig.~\figref{flat.cases}.
\end{pf}

\begin{table}[htbp]
\begin{center}
\begin{tabular}{| c | c | c | c | c | c | l |}
        \hline
$n$ & $a_1$ & $b_0$ & $a_2$ & $b_1$ & $N$ & description
        \\ \hline \hline
any & $0$ & $2$
        & $0$
        & $n$
        & $n+2$
        & pita polyhedra
        \\ \hline
\rule[-6pt]{0pt}{17pt}even & $0$ & $2$
        & $\frac{n}{2}$
        & $0$
        & $\frac{n}{2}+2$
        & flat half-$n$-gon
        \\ 
\rule[-6pt]{0pt}{17pt}even & $2$ & $0$
        & $\frac{n}{2}-1$
        & $0$
        & $\frac{n}{2}+1$
        & flat half-$n$-gon
        \\ 
\rule[-6pt]{0pt}{17pt}odd & $1$ & $1$
        & $\frac{n-1}{2}$
        & $0$
        & $\frac{n+1}{2}+1$
        & flat half-$n$-gon
        \\ \hline
\end{tabular}
\end{center}
\caption{Fold cases for regular $n$-gons, $n > 6$. 
$N$ is the number of polytope vertices.}
\tablab{flat.cases}
\end{table}

\begin{lemma}
For $n \le 6$, regular polygons fold to additional
polytopes (beyond those listed in Lemma~\lemref{ngt6})
as detailed in Table~\tabref{nle6}.
\lemlab{nle6}
\end{lemma}
\begin{pf}
Lemma~\lemref{cut.tree}
limits the possible nonpath cut trees to
`{\tt Y}', `{\tt +}', and `{\tt I}'.
We first argue that
`{\tt I}' is only possible for $n=6$.
The two interior nodes of the tree must have curvatures
in $\{ \a_3, \b_2 \}$.
For $n=3$, there are not enough vertices to make these nodes.
For $n=4$, there are enough vertices to make two $\b_2$
nodes, but this then forces the `{\tt +}' structure,
i.e., the interior edge of the `{\tt I}' has length zero.
For $n=5$ and $n=6$, $\b_2$ is not possible.
For $n=5$, there are not enough vertices to make two $\a_3$ vertices.
And finally, for $n=6$, there are enough vertices, and
the folding produces a flat rectangle.

Thus only `{\tt Y}' and `{\tt +}' are possible.
The `{\tt +}' can only be realized in two ways:
by gluing four vertices together, which is only possible for $n=4$
(see $\a_4$ column in Table~\tabref{alpha}),
and by gluing three vertices to an edge, which is only
possible for $n=3$ (see $\b_3$ column in Table~\tabref{beta}).

There are a number of ways to realize `{\tt Y}'-trees.
The constraint that the curvature add to $4\pi$, Eq.~\eqref{4pi},
together with the discrete set of possible curvatures in
Tables~\tabref{alpha} and~\tabref{beta},
lead to the possibilities listed in Table~\tabref{nle6}.
(The second line of the table
was previously illustrated in Fig.~\figref{tri.tetra}.)
\end{pf}

\begin{table}[htbp]
\begin{center}
\begin{tabular}{| c | c | l | c | c | c | c | l |}
        \hline
$n$ & {\em Tree} & {\em Gluing Description} & {\em Curvatures} 
        & $N$ & {\em Polytope Description}
        \\ \hline \hline
$3$
        & `{\tt Y}'
        & 3 v
        & $\a_3 + 3 \b_0$
        & $4$
        & tetrahedron
        \\ \hline
$3$
        & `{\tt Y}'
        & 2 v + inc e
        & $\a_1 + 2 \b_0 + \b_2$
        & $4$
        & $\infty$ tetrahedra
        \\ \hline
$3$
        & `{\tt Y}'
        & 2 v + adj e
        & $3 \b_0 + 3 \b_1$
        & $4$
        & $\infty$ $5$v polytopes
        \\ \hline
$3$
        & `{\tt +}'
        & 3 v + e
        & $4 \b_0 + \b_3$
        & $4$
        & $\infty$ tetrahedra
        \\ \hline \hline
$4$
        & `{\tt Y}'
        & 3 v
        & $\a_1 + 2 \b_0 + \a_3$
        & $4$
        & tetrahedron
        \\ \hline
$4$
        & `{\tt Y}'
        & 2 adj v + opp e
        & $3 \b_0 +2 \b_1 + \b_2$
        & $5$
        & $\infty$ $5$v polytopes
        \\ \hline
$4$
        & `{\tt Y}'
        & 2 adj v + inc e
        & $4 \b_0 + 2\b_2$
        & $4$
        & $\infty$ tetrahedra
        \\ \hline
$4$
        & `{\tt Y}'
        & 2 adj v + adj e
        & $3 \b_0 +2 \b_1 + \b_2$
        & $5$
        & $\infty$ $5$v polytopes
        \\ \hline
$4$
        & `{\tt Y}'
        & 2 opp v
        & $2 \b_0 + 2 \b_1 + \b_2$
        & $4$
        & $\infty$ tetrahedra
        \\ \hline
$4$
        & `{\tt +}'
        & 4 v
        & $\a_4 + 4 \b_0$
        & $4$
        & flat square
        \\ \hline \hline
$5$
        & `{\tt Y}'
        & 2 adj v + 1 opp v
        & $2 \a_1 + \a_3 + \b_0$
        & $4$
        & tetrahedron
        \\ \hline
$5$
        & `{\tt Y}'
        & 3 adj v
        & $\a_2 + \a_3 + 3\b_0$
        & $5$
        & $5$v polytope
        \\ \hline \hline
$6$
        & `{\tt Y}'
        & 3 alt v
        & $3 \a_1 + \a_3$
        & $3$
        & flat triangle
        \\ \hline
$6$
        & `{\tt Y}'
        & 3 adj v
        & $\a_1 + \a_2 + \a_3 + 2 \b_0$
        & $4$
        & tetrahedron
        \\ \hline
$6$
        & `{\tt Y}'
        & 2 adj v + v
        & $\a_1 + \a_2 + \a_3 + 2 \b_0$
        & $4$
        & tetrahedron
        \\ \hline
$6$
        & `{\tt I}'
        & 3 adj v, 3 adj v
        & $2 \a_3 + 4 \b_0$
        & $4$
        & flat rectangle
        \\ \hline        
\end{tabular}
\end{center}
\caption{Additional fold possibilities for regular $n$-gons, $n \le 6$. 
$N$ is the number of polytope vertices.
Notation in {\em Gluing Description\/} column:
v = vertex,
e = edge,
adj = adjacent,
alt = alternate,
opp = opposite,
inc = included.
In the {\em Polytope Description\/} column:
$\infty$ = continuum of,
$5$v polytope = $5$-vertex polytope.
Each entry of the {\em Curvatures\/} column satisfies 
Eqs.~\eqref{4pi} and~\eqref{indices}.
}
\tablab{nle6}
\end{table}

If we treat the $n$ vertices of a regular $n$-gon as assigned
the same label (as seems appropriate),
Lemmas~\lemref{ngt6} and~\lemref{nle6}
together show that there are only $O(1)$ ways to fold up
a regular polygon, justifying the entry in Table~\tabref{results}.
If we label the vertices with distinct labels, then there are $O(n)$
foldings.

\subsection{Pita Polytopes}
\seclab{pita}
We define a {\em pita polytope\/} as one obtained by
a perimeter-halving folding of a regular polygon at a
point on an edge that is not a midpoint,
as per the first line of Table~\tabref{flat.cases}.
Let the regular $n$-gon $P$ have unit edge length, and let
the fold points $x$ be distance $a$ from $v_0$ along edge $v_0 v_1$.
Let $b = 1-2a$.  Call the point along $\bP$ to which $v_i$ glues
$v'_i$.
See
Fig.~\figref{duodec} for an example with $n=12$.  We will use this
example throughout the section.
\begin{figure}[htbp]
\centering
\includegraphics[width=0.9\textwidth]{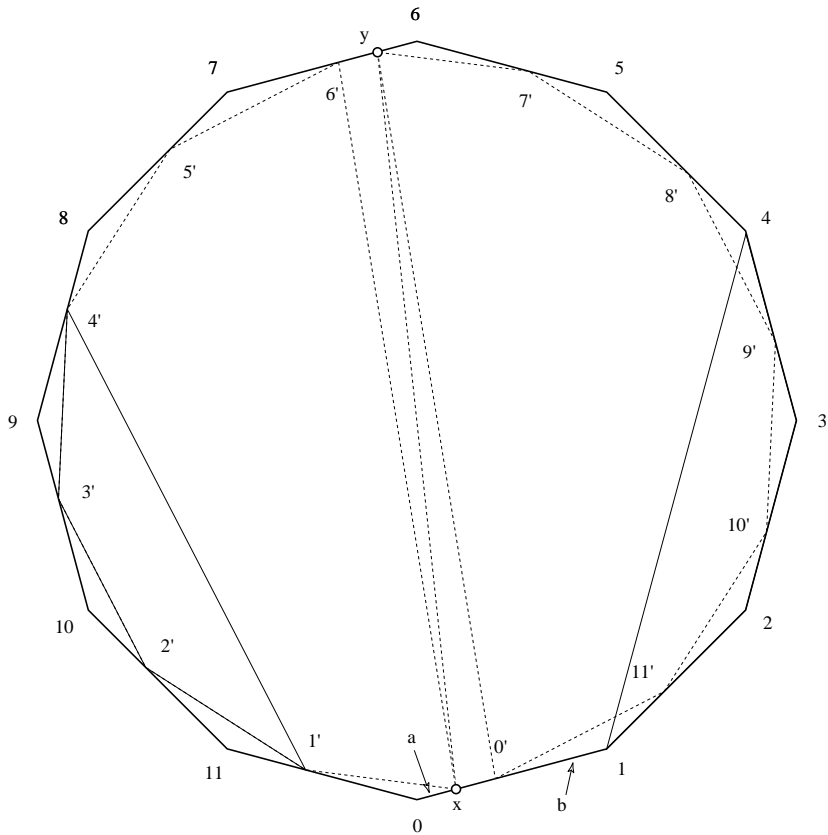}
\caption{Duodecagon, $n=12$, $\a = 30^\circ$,
$a \approx 0.2$, $b \approx 0.6$.
$x$ and $y$ are perimeter-halving fold vertices.
The dashed lines are (largely) conjectured creases.
The vertices $v_i$, and the gluing points $v'_i$,
are labeled with $i$ and $i'$ respectively.
The left and right quadrilaterals play a role
in Fig.~\figref{primes.shorter} below.
}
\figlab{duodec}
\end{figure}

As mentioned in Section~\secref{Introduction},
we have no method for computing the 3D structure of the
unique polytope determined by a particular Aleksandrov gluing.
Moreover, we do not even have a general method for computing the creases,
i.e., the edges of the polytope.
We will therefore largely conjecture the structure of the pita
polytopes in this section, although we will establish
a subset of the creases.
We will only explore the situation for even $n$.
Let $\a = 2\pi/n$, the turn angle at each vertex of the polygon.

We view each pita polytope as composed of four parts:\footnote{
        The relationship to the example in Fig.~\figref{sperm}
        should be evident.
}
\begin{enumerate}
\item A central parallelogram with short side $a$:
$(x, v'_0, y, v'_{n/2})$.
\item A top, nearly half-$n$-gon:
$(v'_0, v'_{n-1}, v'_{n-2}, \ldots, v'_{n/2+1})$.
\item A bottom, nearly half-$n$-gon, congruent by reflection to the top:
$(x, v'_1, v'_2, \ldots, v'_{n/2})$.
\item A ``mouth,'' a strip of triangular teeth;
see Fig.~\figref{strip}.
$n-2$ of the triangles in the strip are
congruent; call their generic shape $T_1$.
$T_1$ has sides of length
$b$, $2a$, and $1$, with an angle $\a$ between $b$ and $2a$.
The two extreme triangles of the mouth are smaller,
of shape $T_2$:
lengths $b$ and $a$ surrounding an angle $\a$.
\end{enumerate}
\begin{figure}[htbp]
\centering
\includegraphics[width=\textwidth]{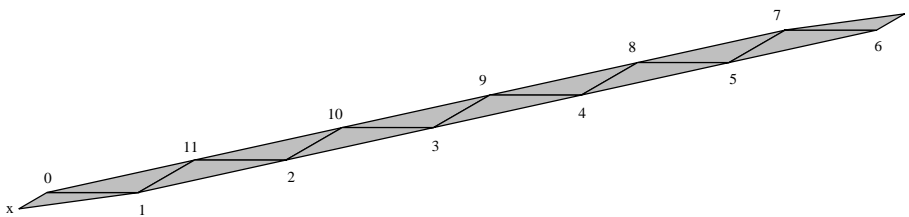}
\caption{Mouth strip of teeth corresponding to Fig.~\figref{duodec}.
(Not to same scale.)
}
\figlab{strip}
\end{figure}
We conjecture that the central parallelogram's edges are
creases, as is its central perimeter-splitting diagonal
$xy$.
Call the top and bottom nearly half-$n$-gons
{\em pita polygons.}
We have no conjectures about how the pita polygons
are triangulated (except that they are triangulated the same).
Finally, we prove below in
Lemma~\lemref{mouth} that the mouth
is creased at the edges displayed in
Fig.~\figref{strip}.

The final 3D shape looks something like
Fig.~\figref{pita}.
As $n \rightarrow \infty$, the polytope approaches a 
doubly-covered flat semicircle.
\begin{figure}[htbp]
\centering
\begin{minipage}{0.41\linewidth}
\centering
\includegraphics[width=\linewidth]{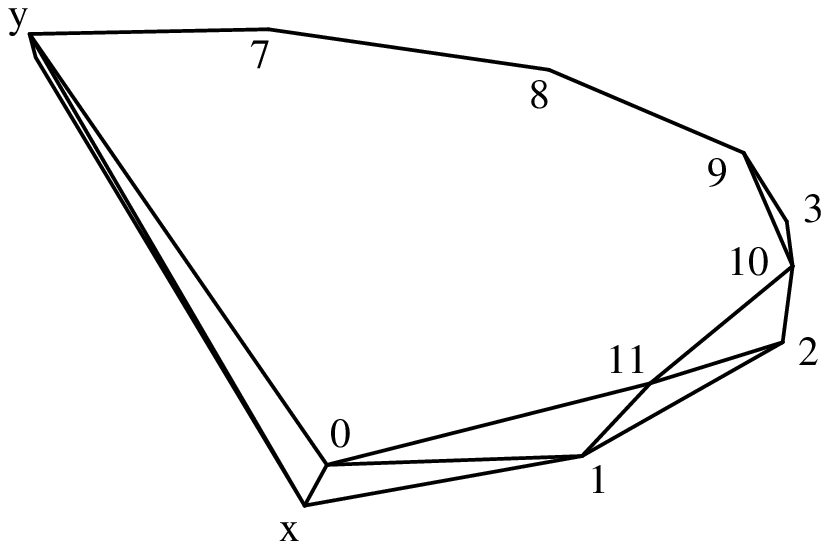}
\end{minipage}%
\hspace{5mm}%
\begin{minipage}{0.49\linewidth}
\centering
\includegraphics[width=\linewidth]{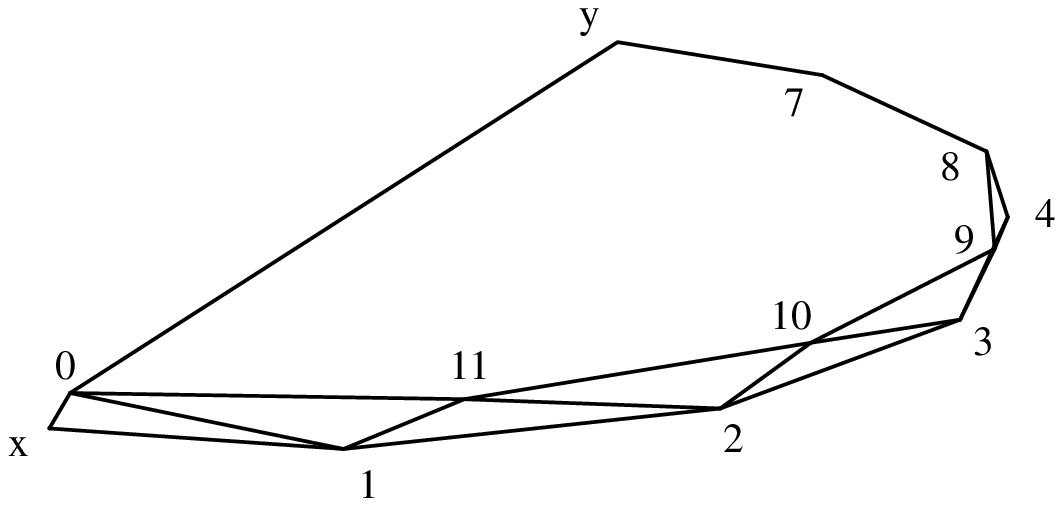}
\end{minipage}
\caption{Two views of the approximate 3D shape of pita polytope folded as per
Fig.~\figref{duodec}.}
\figlab{pita}
\end{figure}

We now establish the structure of the mouth of pita polytopes.
We start with this obvious claim:
\begin{lemma}
Pita polytopes are not flat.
\lemlab{not.flat}
\end{lemma}
\begin{pf}
A flat polytope is a pasting of two congruent polygons,
oriented and aligned the same.  The vertices of the polygons
are the only spots on the polytope surface with curvature.
We know the location of all these $n+2$ vertices:
$x$, $y$, and $v_0,\ldots,v_{n-1}$.
Thus the two polygons must be 
$(y,x,v_1,\ldots,v_{n/2})$
$(x,y,v_{n/2+1},\ldots,v_0)$.
However, because $x$ and $y$ are not at the midpoints of their
edges (by definition of a pita polytope),
these two polygons are not congruent.
\end{pf}

Our tools will be two facts about edges of triangulated polytopes,
neither of which we will prove:

\begin{fact}
Every edge of a polytope is a shortest path between its
endpoint vertices.
\factlab{sp.edge}
\end{fact}

Call two polytope edges incident to the same polytope vertex $v$
{\em adjacent\/} if they are consecutive in a circular
sorting around $v$.
\begin{fact}
The smaller surface angle between two adjacent edges incident to
a polytope vertex is less than $\pi$.
In other words,
within every open semicircle of face angle at a polytope vertex $v$,
there is at least one edge incident to $v$.
\factlab{semi}
\end{fact}

We use Fact~\factref{sp.edge} to eliminate certain geodesics as
candidates for polytope edges.
The following lemma gathers together some basic distance relationships
to be used later to show that some geodesics are not shortest paths:

\begin{lemma}
The following distance relationships hold 
for the length of chords between points
of a pita polygon:
\begin{enumerate}
\item $| v_i - v'_j | = | v'_i - v_j |$.
\item $| v'_i - v'_j | < | v_i - v_j |$ for all $|i-j| > 1$,
i.e., for all $j \neq i$ and $j \neq i \pm 1$.
\item $| v'_i - x | < | v_i - x |$ for all $i \neq 0$.
\item $| v'_i - y | < | v_i - y |$ for all $i \neq n/2$.
\end{enumerate}
\lemlab{distances}
\end{lemma}
\begin{pf}
\begin{enumerate}
\item The polygons cut off by the chords $(v_i,v'_j)$
and $(v'_i,v_j)$ are congruent.
For example, in Fig.~\figref{duodec},
the chord $(v_4,v'_0)$ cuts off a polygon of edge
lengths $(b,1,1,1)$, and
the chord $(v'_4,v_0)$ cuts off a polygon of
lengths $(b,1,1,1)$, both of whose outer interior angles are
all $\a$.
\item
Distances between the $v'_i$ vertices are in general
less than distances between the corresponding unprimed vertices,
because the primed vertices form a regular figure inscribed in
the $n$-gon.
A particular instance is illustrated in
Fig.~\figref{primes.shorter}.  For $j=i+1$, the distances are
equal.
\begin{figure}[htbp]
\centering
\includegraphics[height=6cm]{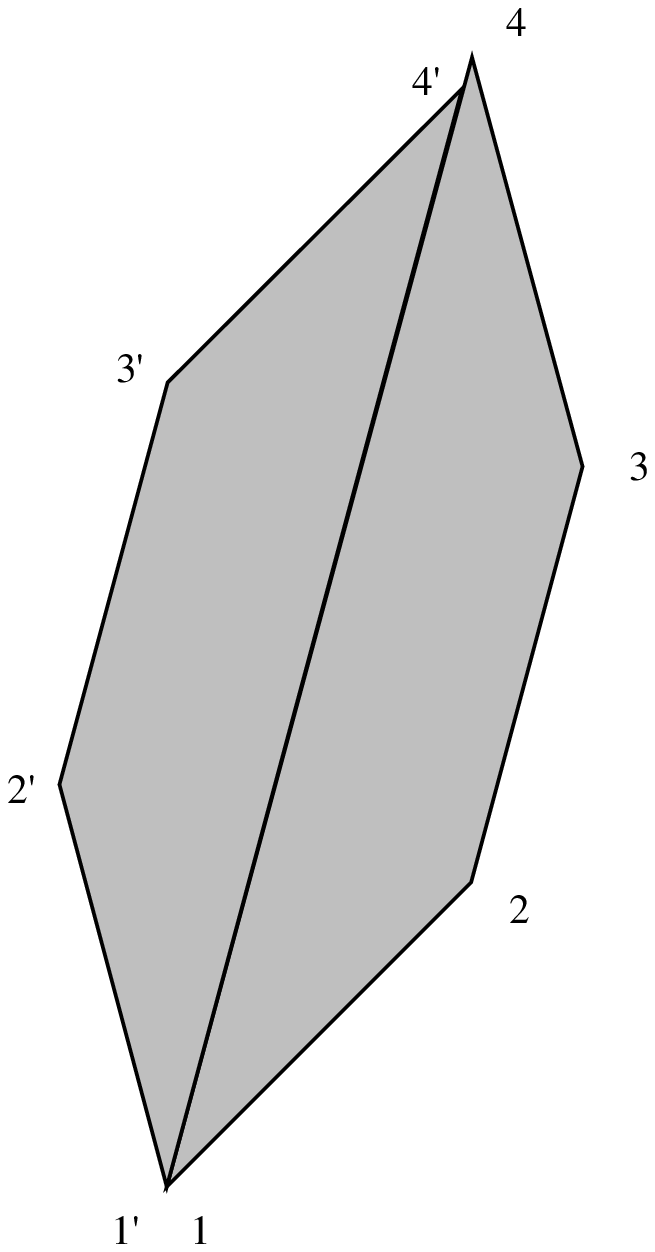}
\caption{$| v'_1 - v'_4 | < | v_1 - v_4 |$ (cf.~Fig.~\figref{duodec}).}
\figlab{primes.shorter}
\end{figure}
\item Here the reason is similar: the primed vertices are
inscribed in the $n$-gon determined by the unprimed vertices.
For example, $|v_1 - x| = a+b$, but
$|v'_1 -x|$ is the length of the hypotenuse of a $T_2$ triangle,
with sides $a$ and $b$, which is shorter by the triangle
inequality.
We will not detail the computations necessary to establish this
claim for all $i$.\complaint{I'm tired}
The only exception to the inequality is for $v_0$, when
$|v_0 - x| = |v'_0 -x| = a$.
\item Symmetric with previous case.
\end{enumerate}
\end{pf}

To eliminate the equal-length geodesics in
Lemma~\lemref{distances}(1),
we will need the following:
\begin{lemma}
An edge $e=vu$ of a nonflat polytope $Q$ is a uniquely shortest path,
i.e., there is not another geodesic of the same length from $v$
to $u$.
\lemlab{uniq.sp}
\end{lemma}
\begin{pf}
Suppose $e=vu$ is an edge of $Q$.
Let $g$ be another geodesic between $v$ and $u$ of the same
length as $e$.  Then because $e$ is a straight segment in 3D,
and because any nonstraight path is strictly longer, it must
be that $g$ is also a straight segment in 3D.  Thus it must
be coincident with $e$.  If $e$ and $g$ are nevertheless distinct,
then they must be on opposite sides of a flat surface.
But then $Q$ must be flat (``pinched'') at $e=g$, 
which by convexity implies
that $Q$ is entirely flat.  This contradicts the assumption of the lemma.
\end{pf}

We now have assembled enough information to pin down the structure
of the mouth:
\begin{lemma}
The mouth of a pita polytope is
triangulated as in Fig.~\figref{strip}:
the edges 
$$(x,v_1,\ldots,v_{n/2},y,v_{n/2+1},\ldots,v_{n-1},v_0)$$
surrounding the mouth, and 
the diagonals $(v_i, v_{n-i})$
and $(v_{n-i+1},v_i)$ that delimit its ``teeth'' 
(cf.~Fig.~\figref{strip}),
are all polytope edges.
\lemlab{mouth}
\end{lemma}
\begin{pf}
Let $v_i$, $i \in \{ 1,\ldots,n/2-1 \}$ be a vertex of the
mouth.  
(It may help to think of $v_4$ in Fig.~\figref{duodec} as
a typical $v_i$ in this proof.)
By Fact~\factref{semi}, there must be a polytope
edge $e$ incident to $v_i$ on the top face in the half plane
bounded by the line through $v_{i-1} v_i$.
By Lemma~\lemref{distances}(2), the other endpoint of $e$
cannot be any $v_j$, $|i-j| > 1$, for all those are longer than
$|v'_i - v'_j|$, the length of an alternate geodesic.
So they are not shortest paths, and are
ruled out by Fact~\factref{sp.edge}.
By Lemma~\lemref{distances}(3-4), the other endpoint of $e$
cannot be $x$ or $y$, for we have restricted $i$ so that
$i \neq 0$ and
$i \neq n/2$.
This leaves $v'_j$ as a possible endpoint of $e$.
But by Lemma~\lemref{distances}(1), $v_i v'_j$ is not
uniquely shortest, which by Lemma~\lemref{uniq.sp} then implies
that $Q$ must be flat, which we know is false by
Lemma~\lemref{not.flat}.

We have excluded all candidates for the endpoint of $e$
except for $j= i \pm 1$.
Because we are examining the semicircle bounded by  $v_{i-1},v_i$,
this leaves $v_{i+1}$ as the only possible endpoint.  Thus
$v_i v_{i+1}$ is an edge of the polytope.

Repeating this argument for the bottom face,
$i \in \{ n/2+1,\ldots,n-1 \}$,
establishes the outer boundary of the mouth,
excluding the edges incident to $x$ and $y$.
Those can be argued similarly.\complaint{I hope}
The teeth diagonals are now easy to see.
We illustrate with $v_4$ in Fig.~\figref{duodec}.
We have just proved that no edge is incident to $v_4$
across the top face. But that top face must be triangulated
somehow.  The only way to triangulate it without using a
diagonal incident to $v_4$ is to include the diagonal
$v'_9 v'_8$.  This means that $v_4 v_8$ and $v_4 v_9$
are edges of the polytope. 
\end{pf}

\bibliographystyle{alpha}
\bibliography{cc}
\end{document}